\documentstyle[12pt,lnfprep,epsfig,epstopdf,amsmath,textcomp,subfig,float]{article}
\newcommand{\be}{\begin{equation}}
\newcommand{\ee}{\end{equation}}
\def\simge{\mathrel{%
      \rlap{\raise 0.511ex \hbox{$>$}}{\lower 0.511ex \hbox{$\sim$}}}}
\def\simle{\mathrel{
      \rlap{\raise 0.511ex \hbox{$<$}}{\lower 0.511ex \hbox{$\sim$}}}}
\newcommand{\Header}{
     \begin{tabular}{rl}
     \hspace{-.3cm}
     \includegraphics{weblogo2.eps} &
       \renewcommand{\arraystretch}{0.5}
       \begin{tabular}{r}
         {\hspace{1cm}~\LARGE\sffamily LABORATORI~NAZIONALI~DI~FRASCATI}\\
         \\
         {\large\sffamily SIDS-Pubblicazioni}\\
       \end{tabular}
       \renewcommand{\arraystretch}{1}
     \end{tabular}
  \vskip 1cm
  \begin{flushright}
  \renewcommand{\arraystretch}{0.5}
    \begin{tabular}{r}
      {\underline{LNF-12-02/2012}}\\    % insert here the preprint number
      {\small April 2, 2012} \\      % insert here the preprint Date
      \\
    \end{tabular}
  \end{flushright}
  \renewcommand{\arraystretch}{1}
  \vskip 1 cm
  }
\begin{document}
\begin{titlepage}
\title{
  \Header
  {\large \bf Diffraction Radiation of Electron Bunches for One- and Two Slit  Systems}
}
\author{
   V. Shpakov, and S.B. Dabagov\\
{\it ${}^{1)}$INFN Laboratori Nazionali di Frascati, Frascati, Italy}\\
{\it ${}^{2)}$RAS P.N. Lebedev Physical Institute \& NRNU MEPhI, Moscow, Russia}
}
\end{titlepage}

\maketitle
\baselineskip=14pt

\begin{abstract}

New research in acceleration physics leads to growing up the power of charged particles bunches. Existed methods based on interaction of detection devices with bunches do not satisfy our need, because of the fact that new high intensity bunches can damage this devices. Moreover, these methods do not allow analyzing the bunches in real time. Recently new technique based on ODRI (optical diffraction radiation interference) by a bunch at its propagation through the slit  was proposed. In this work the results of theoretical simulations on diffraction radiation, in particular ODRI, by electrons for various slit systems as well as the comparison with the DESY experimental data  are presented.
\end{abstract}

\vspace{4cm}

\newpage
\section{Introduction}

As known, a charged particle emits electromagnetic radiation at its acceleration independently on the origin of this acceleration \cite{1}.  This phenomenon well studied both theoretically and experimentally, moreover, for more then 50 years it has been used within various branches of the science, from the basic research (acceleration physics, particle physics, etc.) to many applications as a source of radiation (synchrotron and undulator radiations,  free electron lasers, applications in material sciences, chemistry, biology, medicine, etc.).

However, a charged particle can emit electromagnetic radiation even without its acceleration, in presence of various media, for instance. The pioneering works on this radiation brings us to the discovery by P. Cherenkov and S. Vavilov, presently known as Cherenkov (or Cherenkov-Vavilov) radiation, the origin of which was first suggested within the phenomenology developed by their colleagues I. Tamm and I. Frank \cite{2}.
This radiation is emitted when a charged particle crosses the media with the velocity exceeding the speed of light in media. Later on, developing the proposed theory for charged particles propagation in solids,  I. Tamm and V. Ginzburg predicted another kind of electromagnetic radiation that was called transition radiation \cite{2,3}. The latter is emitted when the particle crosses the media interfaces characterized by different dielectric constants, thereby, creating polarization currents that, in turn, explains the origin of radiation. Obviously, in this case, to suppose that power (intensity)  independently, the possibility of electromagnetic radiation emission at propagation of charged particle bunches near the solid ends that can be also characterized by induced polarization in media. And one of these types of radiation is known as diffraction radiation (DR), and can be registered at propagation of a charged projectile of constant velocity near the sharp solid end \cite{4}. Special interest represent so-called optical diffraction radiation interference (ODRI)\footnote{Below we use a simple definition as DR}, the features of which was previously discussed in \cite{9}. Indeed the sensitivity of this technique based on the interference is higher.

As seen, DR, by its nature, is a kind of transition radiation but with some essential features. Of course, and it is the name, this phenomenon appears when a charged particle moves near optical heterogeneity, hence, it requests taking into account the diffraction effects on the media end. The theory of diffraction radiation was in details considered by many authors (see, for instance \cite{5} and Refs. in), and presently, we can apply it for various schemes either to predict or to analyze the results of both planned and performed experiments.

In this work we present the results of theoretical analysis for various possible schemes to be experimentally realized. The studies were done to reveal fine features of DR in both near and far zones of radiation for one-slit as well as two-slit systems. The latter induced us to use the wave approximation at the radiation distribution, spatial and angular, and radiation intensity simulations.

\section{Basics of DR on a planar slit}

Below, following the description presented in \cite{5}, we briefly introduce to the basics of diffraction radiation by relativistic electrons on a planar slit.

\begin{figure}[h]
\begin{minipage}{0.57\textwidth}
 \includegraphics[width=0.9\textwidth]{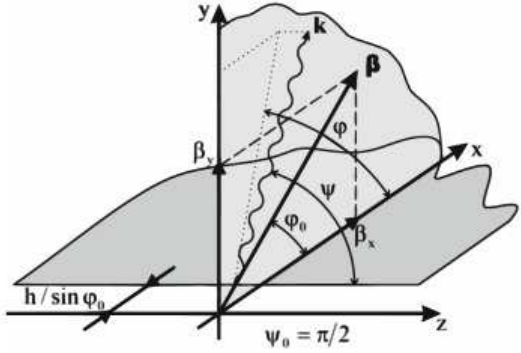}
\end{minipage}
\begin{minipage}{0.37\textwidth}
\vspace{5cm}
  \caption{Geometry of DR on a half-plane (reproduced from \cite{1}).}
  \label{fig1}
\end{minipage}
\vspace{-0.5cm}
\end{figure}

\begin{figure}[h]
\vspace{-1cm}
\begin{minipage}{8cm}
\includegraphics[scale=0.3]{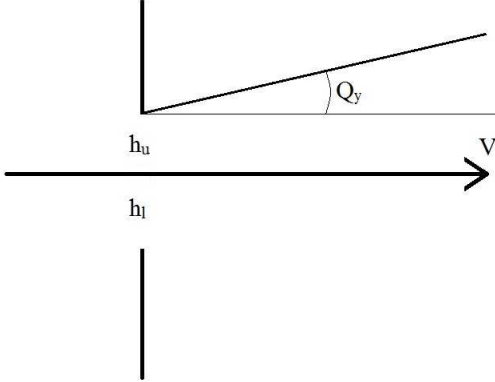}
\end{minipage}
\begin{minipage}{6.5cm}
\vspace{4cm}
\caption{Schematic view of a slit and the emission angle of DR with respect to the projectile velocity.}
\label{fig2}
\end{minipage}
\end{figure}

As known DR is observed  when charged particle moves near the solid media end. Let consider the case of a perfectly conductive media, i.e. an infinite thin screen, the geometry of which is shown in Fig.~\ref{fig1}. Following the method described in \cite{4,5} we can define the components of current and charge density induced by passing particle in such geometry by the following expressions (here $\hbar=c=1$)
\begin{multline}
\rho=-\frac{B_{q\omega}}{\sqrt{\omega{}\sin\psi(1+\cos\varphi)}}\times \\ \times\left\{\frac{1}{\sin\psi}-\frac{\alpha_0\cos\varphi_0+i\gamma^{-2}
 \sin\varphi_0}{\alpha_0\beta}\frac{\sin\psi(1+\cos\varphi)}{-\sin\psi\cos\varphi+\frac{\cos\varphi_0}{\beta}+i\frac{\alpha_0}{\beta}\sin\alpha_0}\right\}
\end{multline}
$$
j_y=0
$$
where the parameters
$$
B_{q\omega}=\frac{e}{4\pi_2}\frac{\beta\sqrt\omega[\sin\psi-\frac{\cos\varphi_0}{\beta}-i\frac{\alpha_0}{\beta}\sin\varphi_0]}{\omega(\alpha_0\cos\varphi_0+
i\sin\varphi_0)}\exp{(-\ae)}\,\, ,
$$
$$
\alpha_0=\sqrt{\gamma^{-2}+\beta^2\cos^2\psi}=\sqrt{1-\beta^2\sin^2\psi}\,\, ,
$$
$$
\ae=\frac{2\pi{}h}{\beta\lambda}\sqrt{1-\beta^2\sin^2\psi}\,\, ,
$$
correspond  to the particle field, the relativistic angular parameter and attenuation parameter respectively.
\noindent The z-component of  current $j_z$ we can reduce from the continuity equation $\mathbf{nj}=\rho$, where $\mathbf k=\mathbf n\omega$, $\beta$ is the velocity, $\gamma$ is the Lorentz factor, $h$ is the impact parameter and $e$ is the charge of electron. The Fourier components of the field we can be reduced directly from the Maxwell's equations \cite{4,5}. For simplicity, the coordinate system associated with the wave vector $\mathbf k$ is selected:

$$
{\mathbf e}_2=\frac{[{\mathbf z}_0{\mathbf n}]}{\sin\psi}=\{-\sin\phi, \cos\phi, 0\}\,\, ,\
$$
$$
{\mathbf e}_1=-[{\mathbf n}{\mathbf e}_2]=\{\cos\phi\cos\psi,\sin\phi\cos\psi,-\sin\psi\}\,\, ,\
$$
$$
{\mathbf n}=\frac{{\mathbf k}}{\omega} \,\,
$$
The coordinate system is connected to the basic through the appropriate angles. In this coordinate system one of the components of the field is always equal to zero, and the component along $\mathbf{e}_1$ could be neglected due to the reason described below. Then the component $E_2$ for the velocity of a particle perpendicular to the screen can be presented as
\begin{multline}
E_2=-\omega\frac{e}{4\pi^2}\frac{\beta\sqrt{\omega(\sin{\psi}-i\frac{\alpha_0}{\beta})}}{i\omega\sqrt{\omega\sin{\psi(1+\cos{\varphi})}}}
\exp{(\ae)}\times\frac{\sin{\psi}+i\frac{\alpha_0}{\beta}}{-\sin{\psi\cos{\varphi}+i\frac{\alpha_0}{\beta}}}\sin{\varphi}
\end{multline}
Below we can use these relations substituting the variables by $\Theta_x=\pi/2-\psi$ and $\Theta_y=\pi/2-\varphi$ that, as seen from the draw of Fig.~\ref{fig1}, allows us to pass our considerations to the angles of radiation with respect to the particle velocity direction\footnote{It is important to underline that we consider the case of relativistic particles} (see Fig.~\ref{fig2}). As known, the radiation of relativistic particles is mostly  concentrated within the cone $\sim\gamma^{-1}$ (where $\gamma$ is the Lorentz-factor). Taking this fact  into account, after some transformations and neglecting terms of the order of $\sim\gamma^{-2}$, the previous expression can be rewritten for DR from one half-plane in more simple form
\begin{equation}
E_2=\frac{e}{4\pi^2}\frac{\exp{[-\frac{2\pi h}{\lambda}\sqrt{\gamma^{-2}+\Theta_{x}^{2}}}]}{\sqrt{\gamma^{-2}+\Theta_x^2}-i\Theta_y} \,\,,
\label{E2-2}
\end{equation}
where $h$ is the impact parameter, and $\lambda$ is the wavelength of observed radiation.

Using the slit instead of a screen we have to take into account the interference of DR from two half screens (planes), which together act as a unique slit. Thus, total field from the slit  can be defined in the form
\begin{equation}
\mathbf E_{slit}=\mathbf E_{u}\exp{(i\varphi_u)}+\mathbf E_{l}\exp{(-i\varphi_l)}\,\,,
\end{equation}
in which the coefficients "$u$" and "$l$" correspond, respectively, to the upper and lower half-planes. The expression for radiation from upper half-plane we can get from the expression for lower half-plane by the complex conjugation combined with simultaneous substitution $h_l\rightarrow h_u$. Within the approximation of small angles in relativistic case it is possible to define that
\begin{equation}
\varphi_u\sim-\frac{2\pi h_u}{\lambda}\Theta_y,\ \ \ \ \varphi_l\sim-\frac{2\pi h_l}{\lambda}\Theta_y
\label{varphi_ul}
\end{equation}

Then, using Eqs.~(\ref{E2-2}) and (\ref{varphi_ul}) the total field of DR from a slit in the forward direction can be written as follows
\begin{equation}
E_2=\frac{e}{4\pi^2}\left(-\frac{\exp{[-\frac{2\pi h_u}{\lambda}(\sqrt{\gamma^{-2}+\Theta_{x}^{2}}}-i\Theta_y)]}{\sqrt{\gamma^{-2}+\Theta_x^2}-i\Theta_y}+\frac{\exp{[-\frac{2\pi h_l}{\lambda}(\sqrt{\gamma^{-2}+\Theta_{x}^{2}}}+i\Theta_y)]}{\sqrt{\gamma^{-2}+\Theta_x^2}+i\Theta_y}\right)
\label{E2}
\end{equation}

\section{Dependence of DR on projectile-slit position}

For many problems in high energy physics, especially at accelerators when the beams of particles move inside various cavities such as the beam-pipes, solenoids, magnets, etc., it is important to know the position of a particle in a space inside these guides. These features can be evaluated studying DR dependence on the position between the beam and the slit.  Let us analyze the DR intensity versus the angle of emission at various positions of a particle with respect to the center of a slit. Fist of all, below we simplify the task in order to determine some parameters.

Assuming that both radiation wave vector and vector of electron velocity are situated in one plane, we get  $\Theta_x=0$. Moreover, the component $E_1$ linearly depends on $\Theta_x$, hence we can take  $E_1=0$. And that is the reason why we could omit this component from our consideration. That, in experiment can be realized by means of the polarizer. Thus, in this approximation Eq.(\ref{E2}) is reduced to
\begin{equation}
E_2=\frac{e}{4\pi^2}\left(-\frac{\exp{[-\frac{2\pi h_u}{\lambda}(\gamma^{-1}-i\Theta_y)]}}{\gamma^{-1}-i\Theta_y}+\frac{\exp{[-\frac{2\pi h_l}{\lambda}(\gamma^{-1}+i\Theta_y)]}}{\gamma^{-1}+i\Theta_y}\right)
\end{equation}

To perform concrete analysis of DR angular distribution we have used the parameters of the  DESY test facility, i.e. the bunch and radiation parameters are, respectively, $\gamma=1840, \lambda=800$ nm, and the width of a slit  is 1 mm. The simulations were performed for various electron positions in the space between two half-planes.

\begin{figure}
\vspace{-2cm}
  \centering
  \subfloat[]{\includegraphics[width=0.42\textwidth]{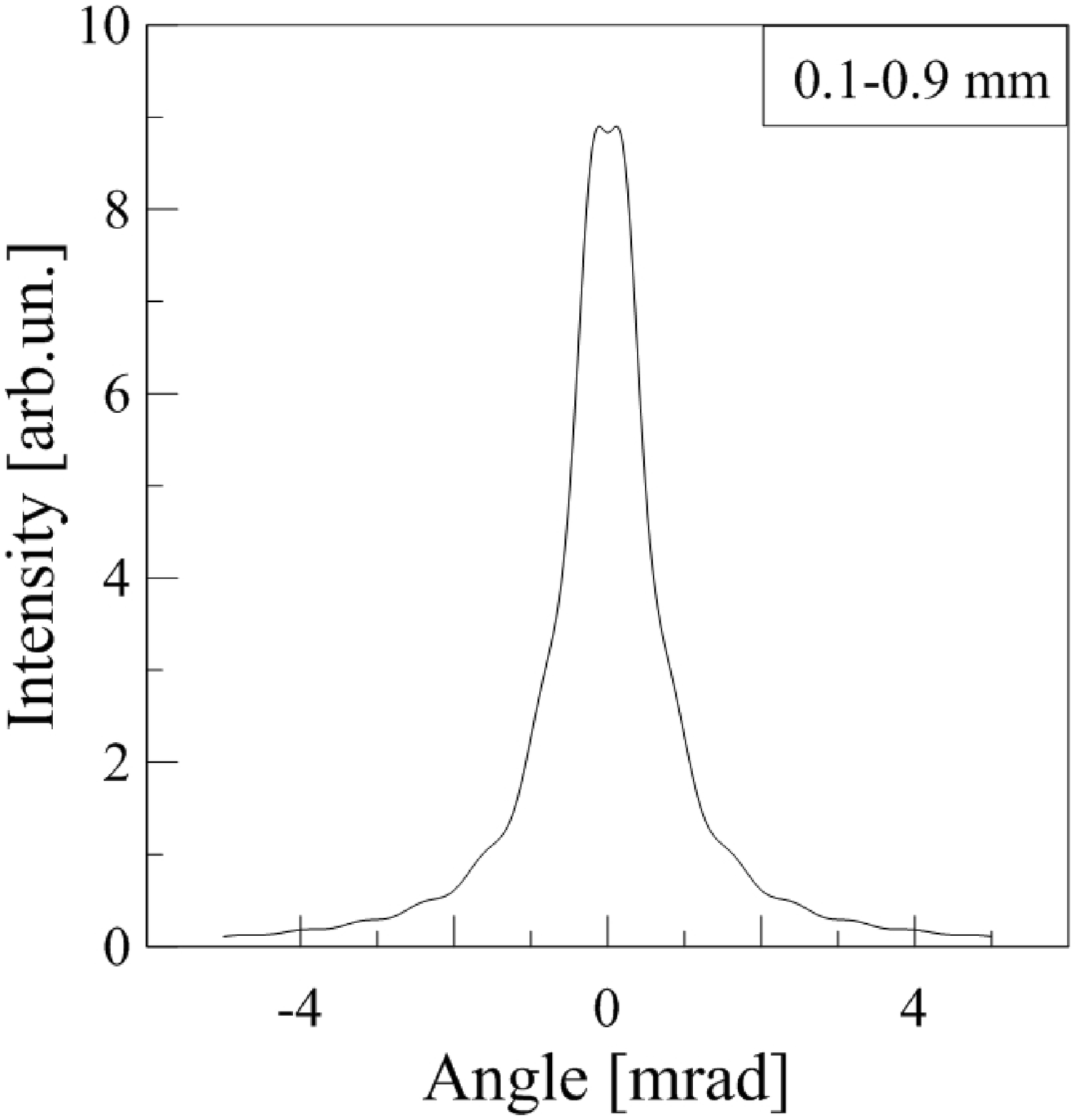}}
           ~
  \subfloat[]{\includegraphics[width=0.42\textwidth]{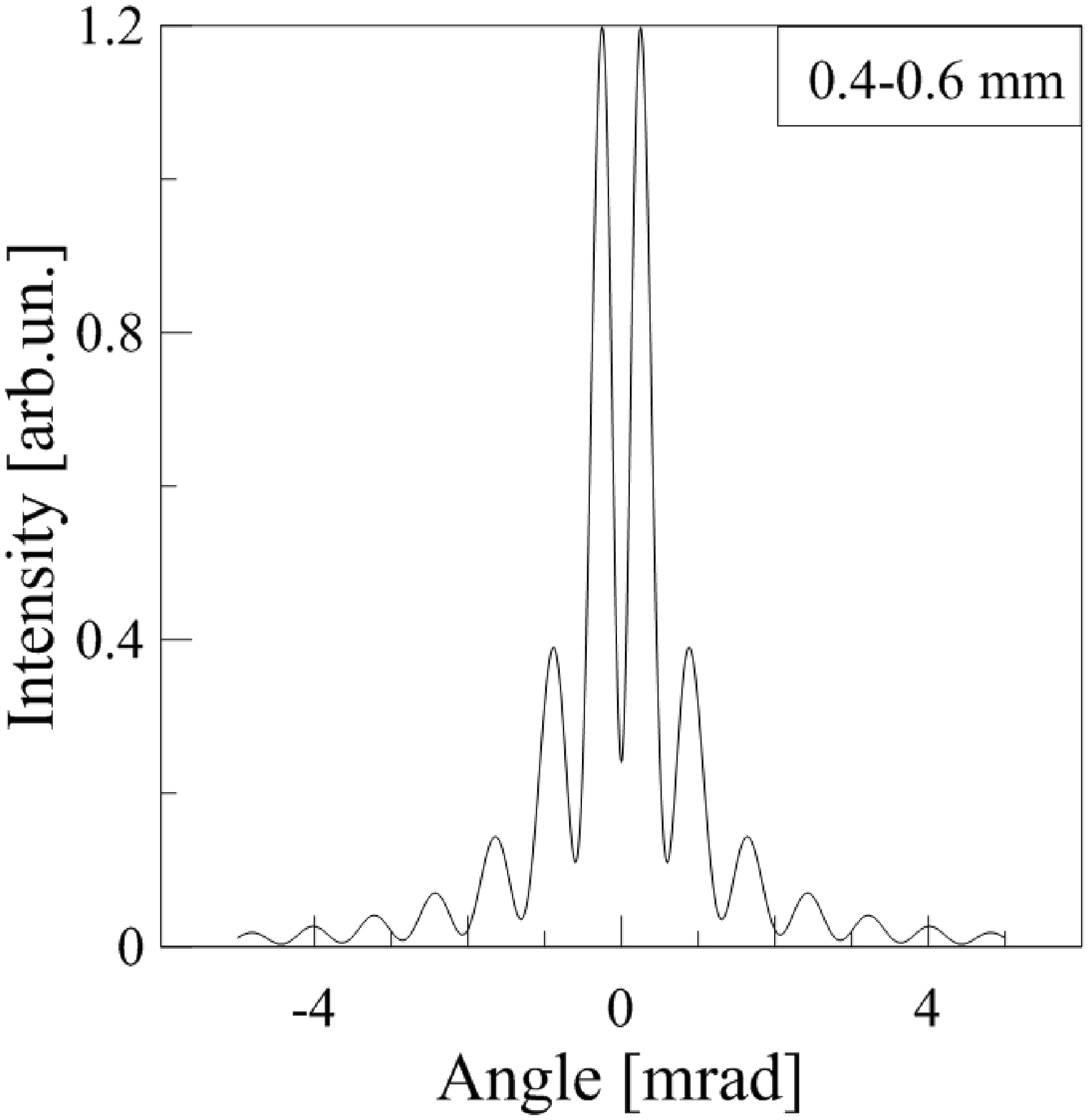}} \\
           ~
  \subfloat[]{\includegraphics[width=0.42\textwidth]{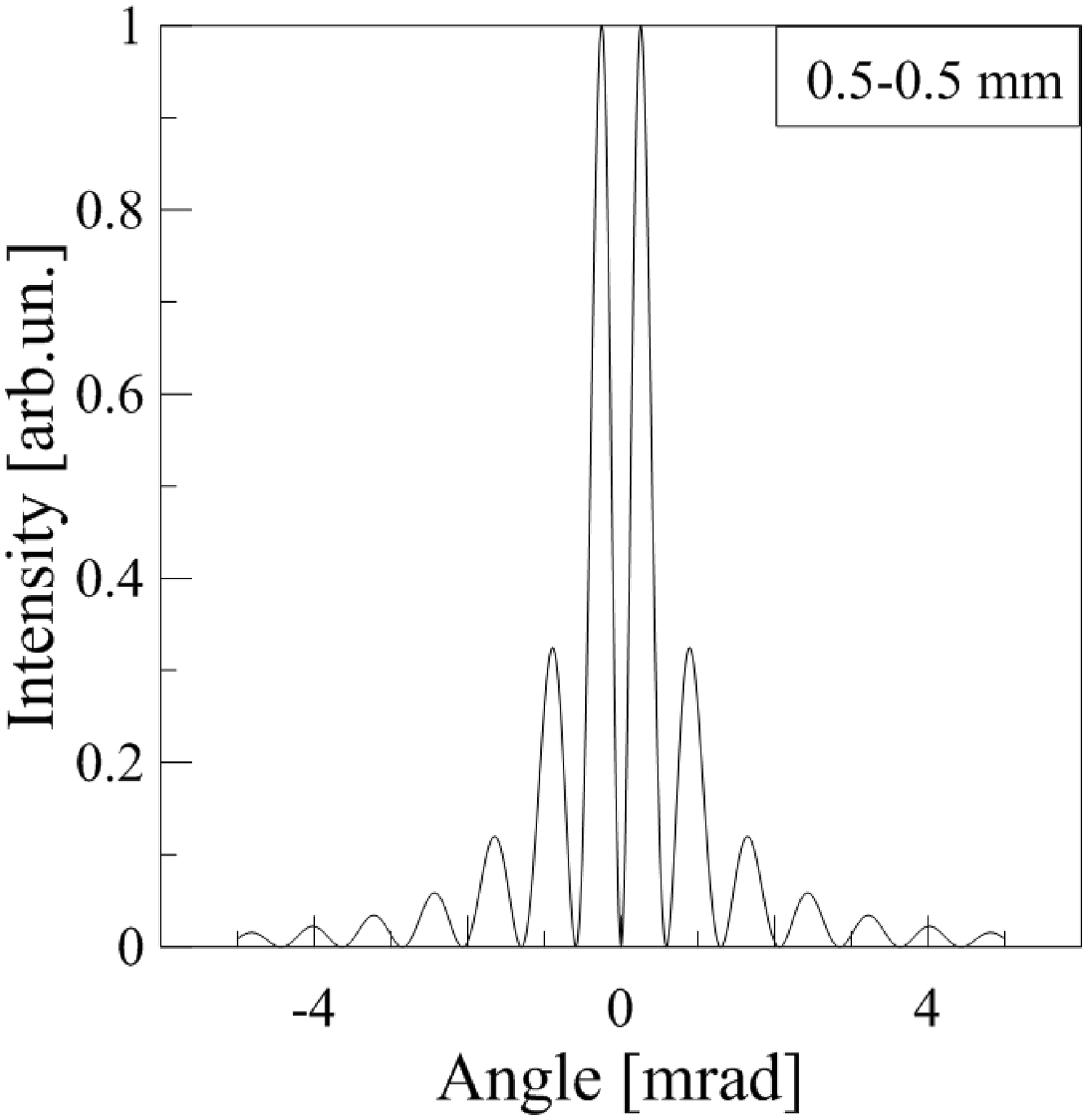}} \\
           ~
  \subfloat[]{\includegraphics[width=0.42\textwidth]{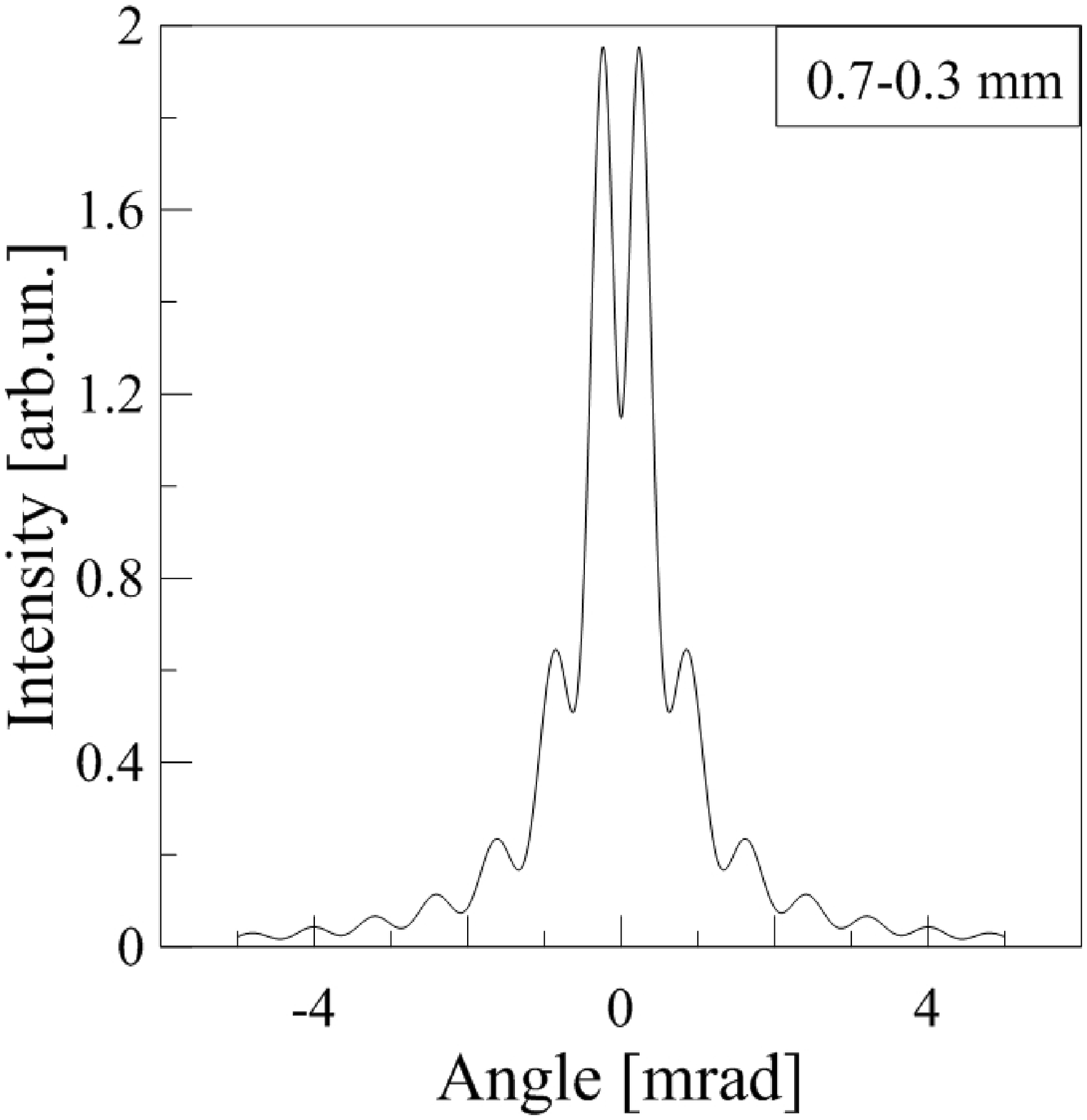}}
           ~
  \subfloat[]{\includegraphics[width=0.42\textwidth]{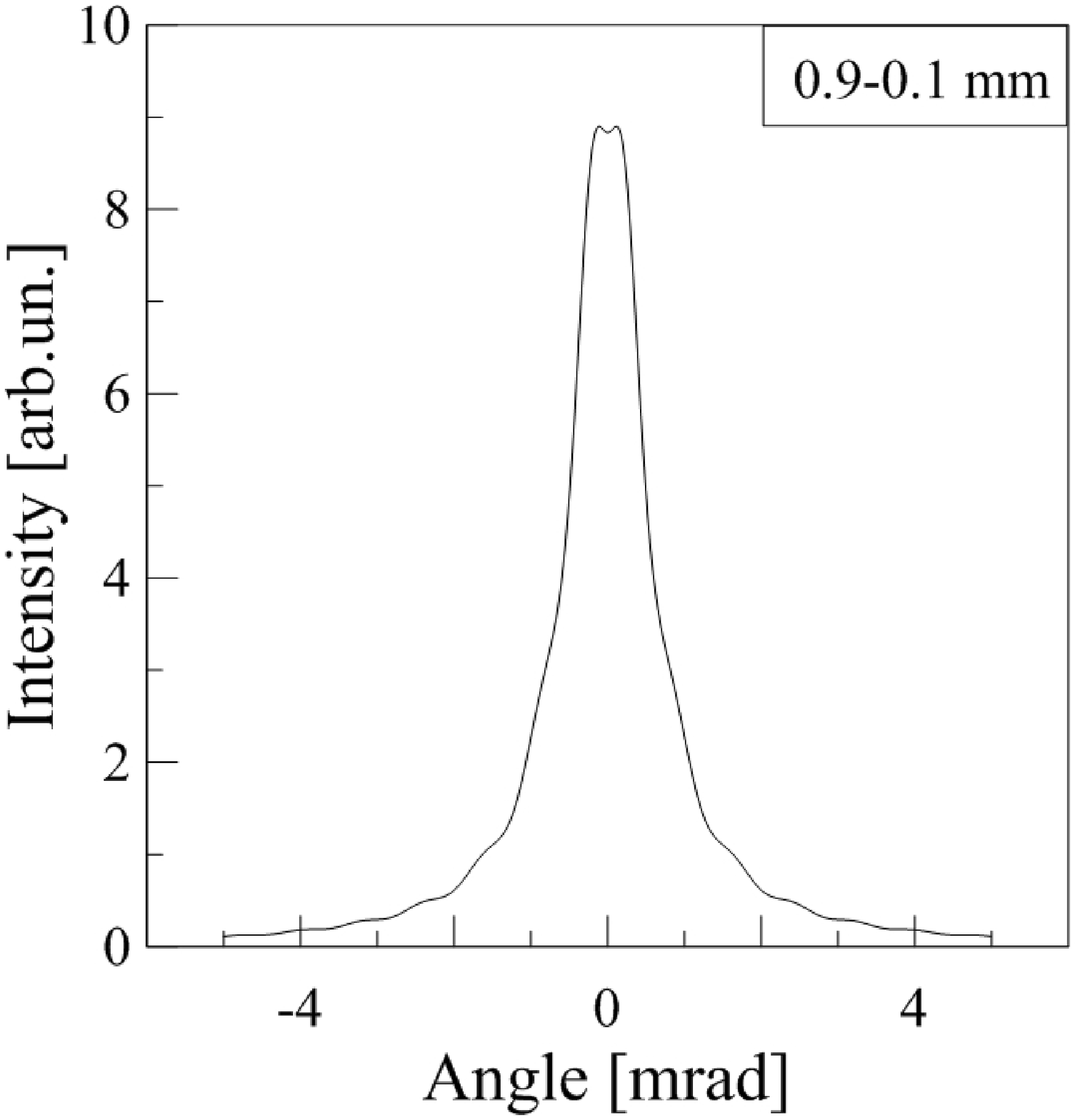}}
           ~
  \caption{DR angular dependence in dependence on the position of a projectile with respect to both half-planes of a slit. The distance between the particle and plane has been marked above each plot, where the first value is the distance to the upper edge, while the second number - to the lower edge.}
  \label{fig3}
\end{figure}

\begin{figure}[h]
\vspace{-2cm}
 \begin{minipage}{0.5\textwidth}
 \vspace{5.5cm}
  \caption{The angle between the spikes vs the position of a particle with respect to the slit center.}
   \label{fig4}
\end{minipage}
\begin{minipage}{0.45\textwidth}
\hfill\includegraphics[scale=0.4]{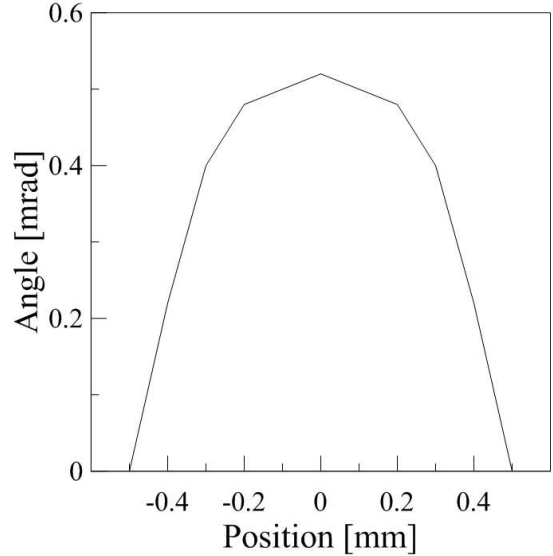}
\end{minipage}
\end{figure}

In Fig.~\ref{fig3} we have presented the angular evolution of DR in dependence on the position of a projectile with respect to both half-planes of a slit. The distance of a particle to the plane has been marked above each plot, the first value is the distance to the upper edge, while the second number - to lower edge. As we can see the distance between the spikes varies at the change of electron position. This dependence is shown in Fig.~\ref{fig4}.

It should be noted that such dependence leaves one uncertainty. We obtain the equal angles between the spikes for two symmetric positions relative to the center of a slit. However, this problem can be resolved if to move the slit detecting simultaneously the change of the angular distance between the spikes. For example, if up-moving the slit we observe the decrease of the angular distance between the spikes, then the projectile is positioned in the upper part of a slit, and vice versa. Obviously, the sensitivity of this method is much higher near one of the edges.

\section{Additional parameters}
In previous section we have considered the point like beam (which corresponds to single projectile) to simplify our task. However, in reality the task is more complex. In order to be closer to the reality we have to take into account a few more parameters. In next sections we consider the size of a beam, its divergence, and, moreover, for the case of two slit system the slits uncoplanarity  will be introduced.

\subsection{Beam size}
Let us start our analysis from the size of a beam. Typically, the beam profile is characterized by the Gaussian distribution in perpendicular cross-section. If $y_0$ is the distance from the arbitrary point of a beam to the center of a slit, we can determine the beam spatial distribution as a function $G(y_0)$. Successfully,total  radiation from the beam can be presented as
\begin{equation}
I_f=\int_a^b \! |E_2(y_0)|^2G(y_0)dy_0,
\end{equation}
where $a$ and $b$ is the limits describing the beam size.

For the Gaussian profile of a  beam we can define a new parameter, a spatial dispersion, which determines the integration limits  $a$ and $b$. Indeed these parameters are equal to the dispersion $\sigma$ \cite{6,7}. Performing some calculations we can reduce the size of a beam with respect to the spikes position in DR angular distribution.
\begin{figure}
\vspace{-2cm}
  \centering
  \subfloat[]{\includegraphics[width=0.45\textwidth]{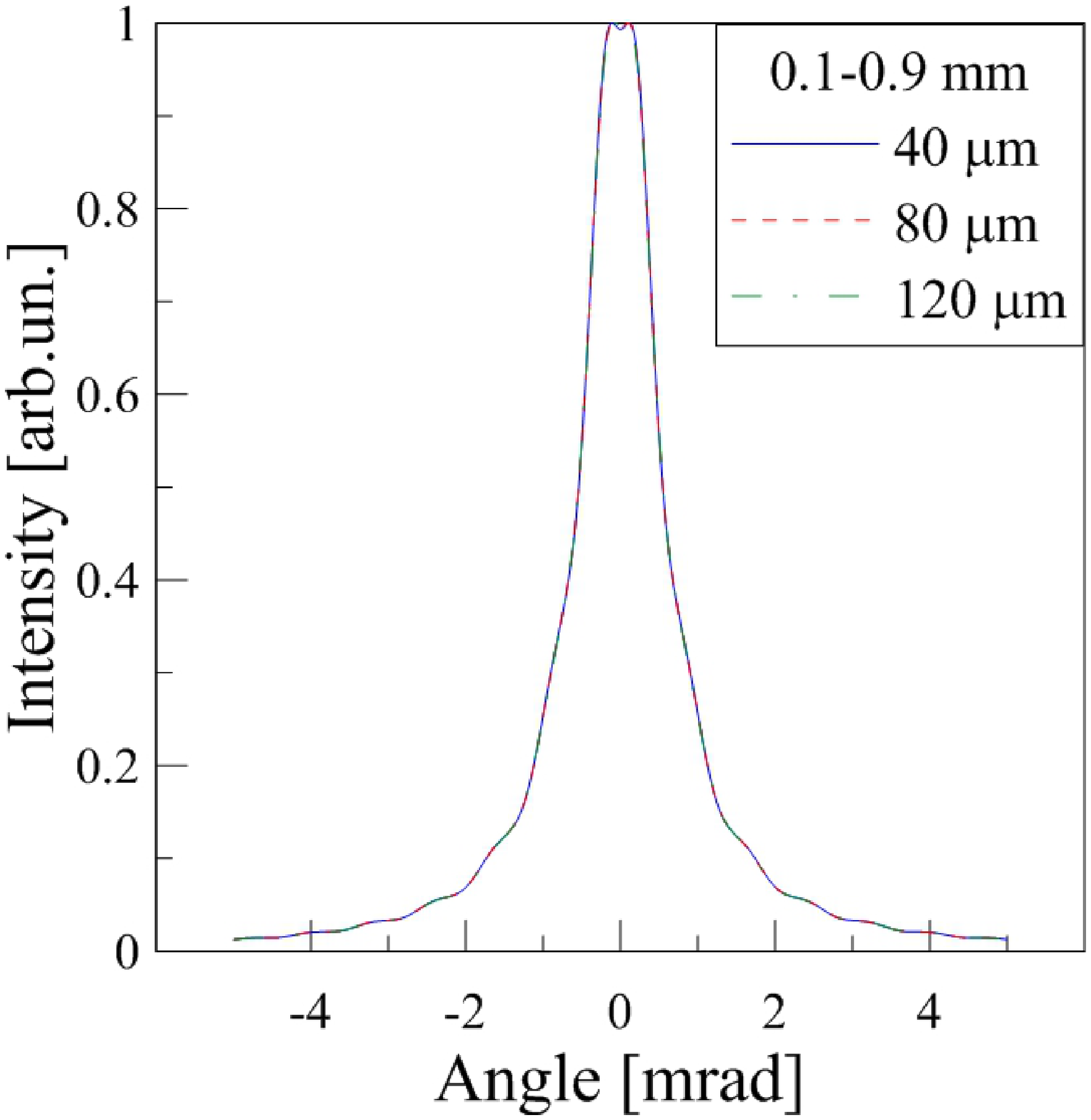}}
             ~
  \subfloat[]{\includegraphics[width=0.45\textwidth]{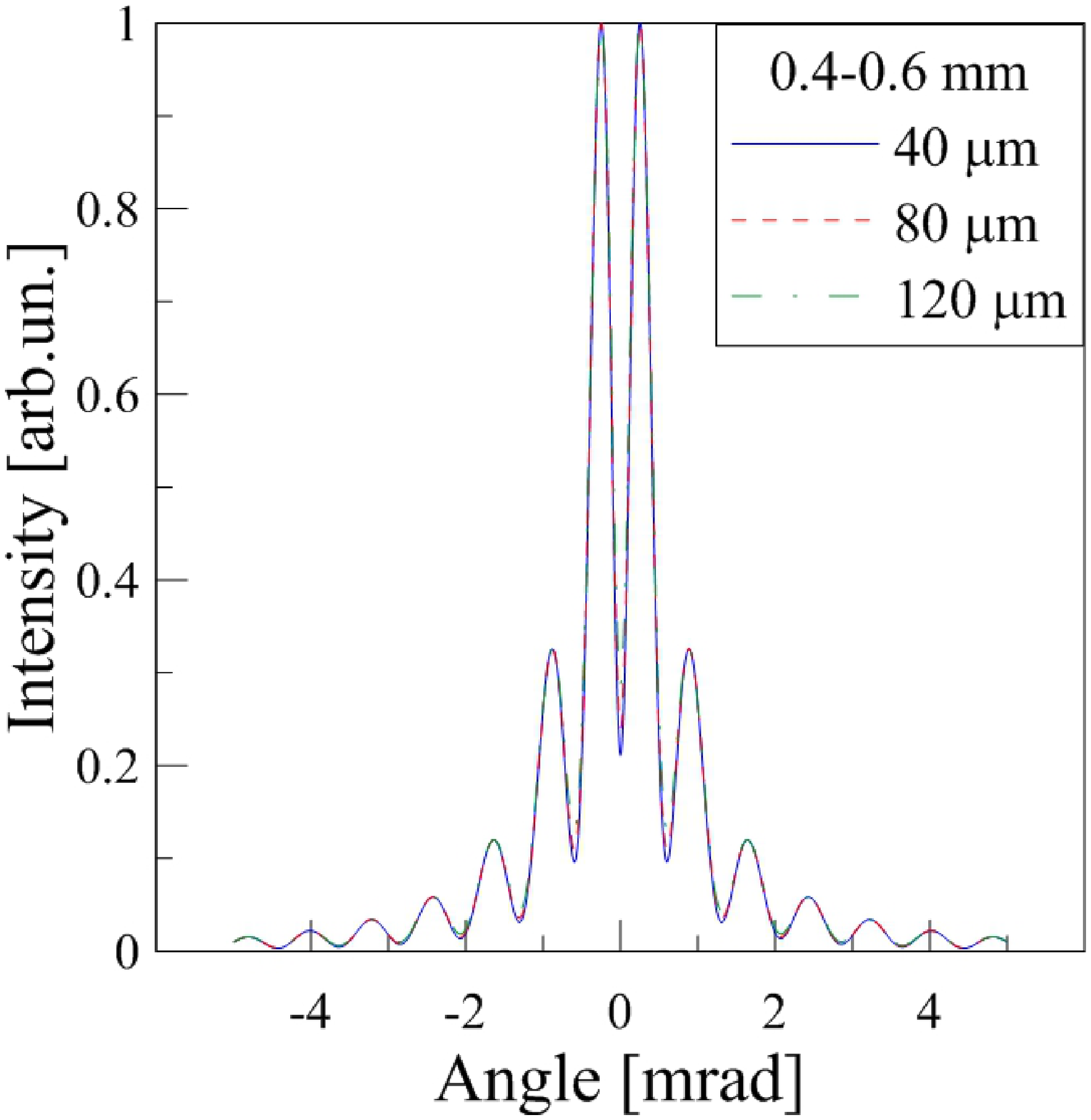}} \\
             ~
  \subfloat[]{\includegraphics[width=0.45\textwidth]{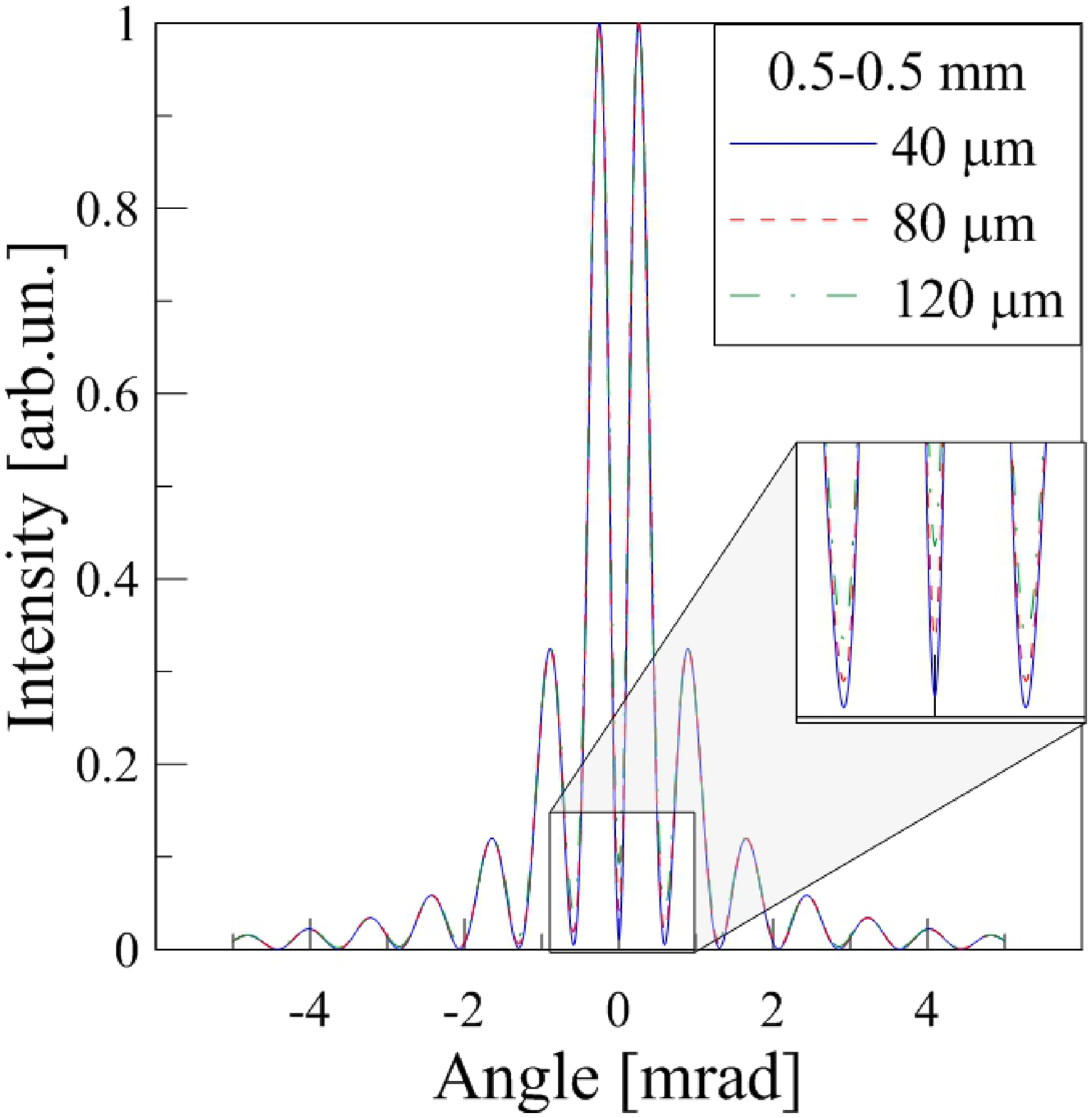}} \\
             ~
  \subfloat[]{\includegraphics[width=0.45\textwidth]{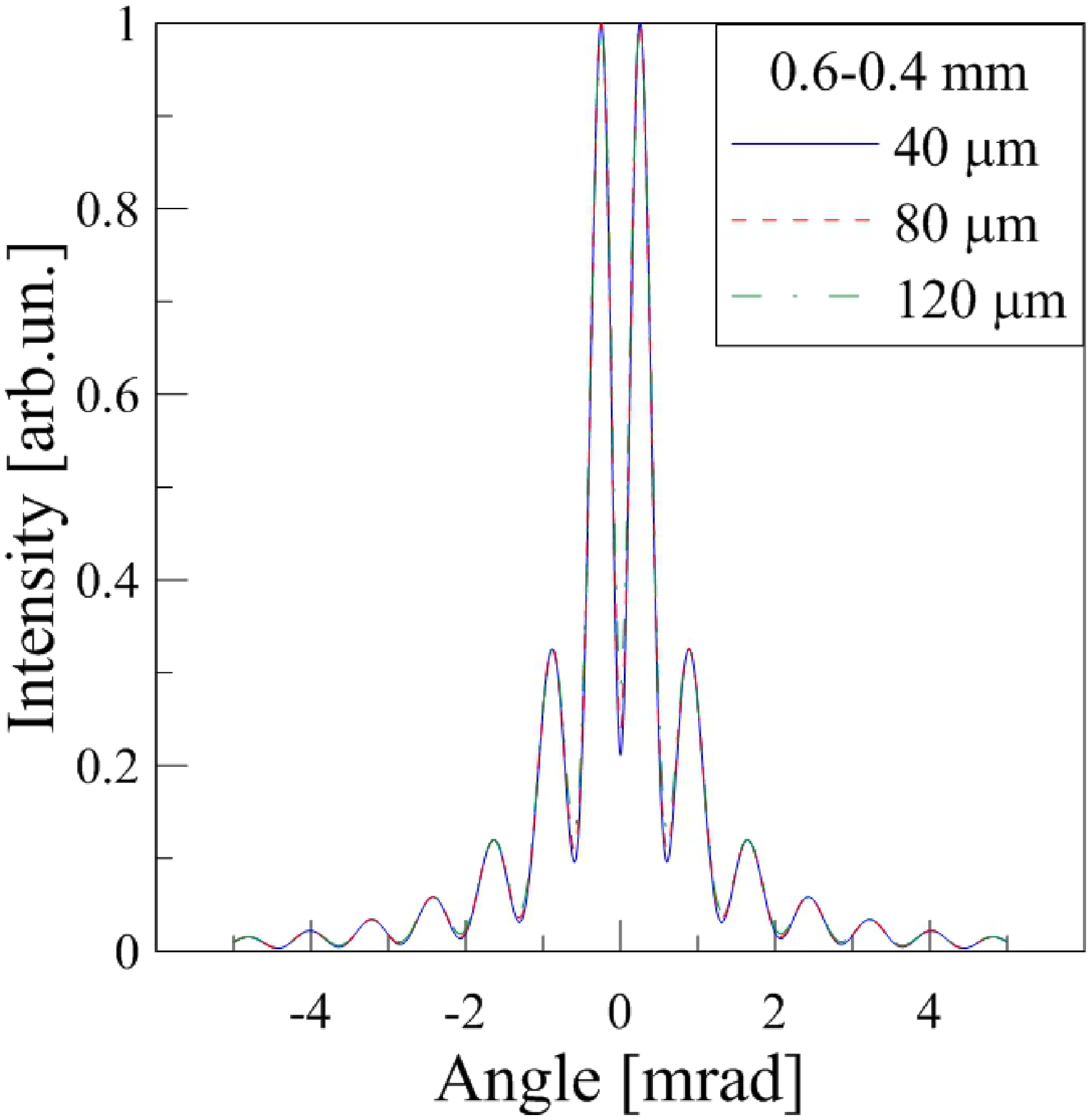}}
             ~
  \subfloat[]{\includegraphics[width=0.45\textwidth]{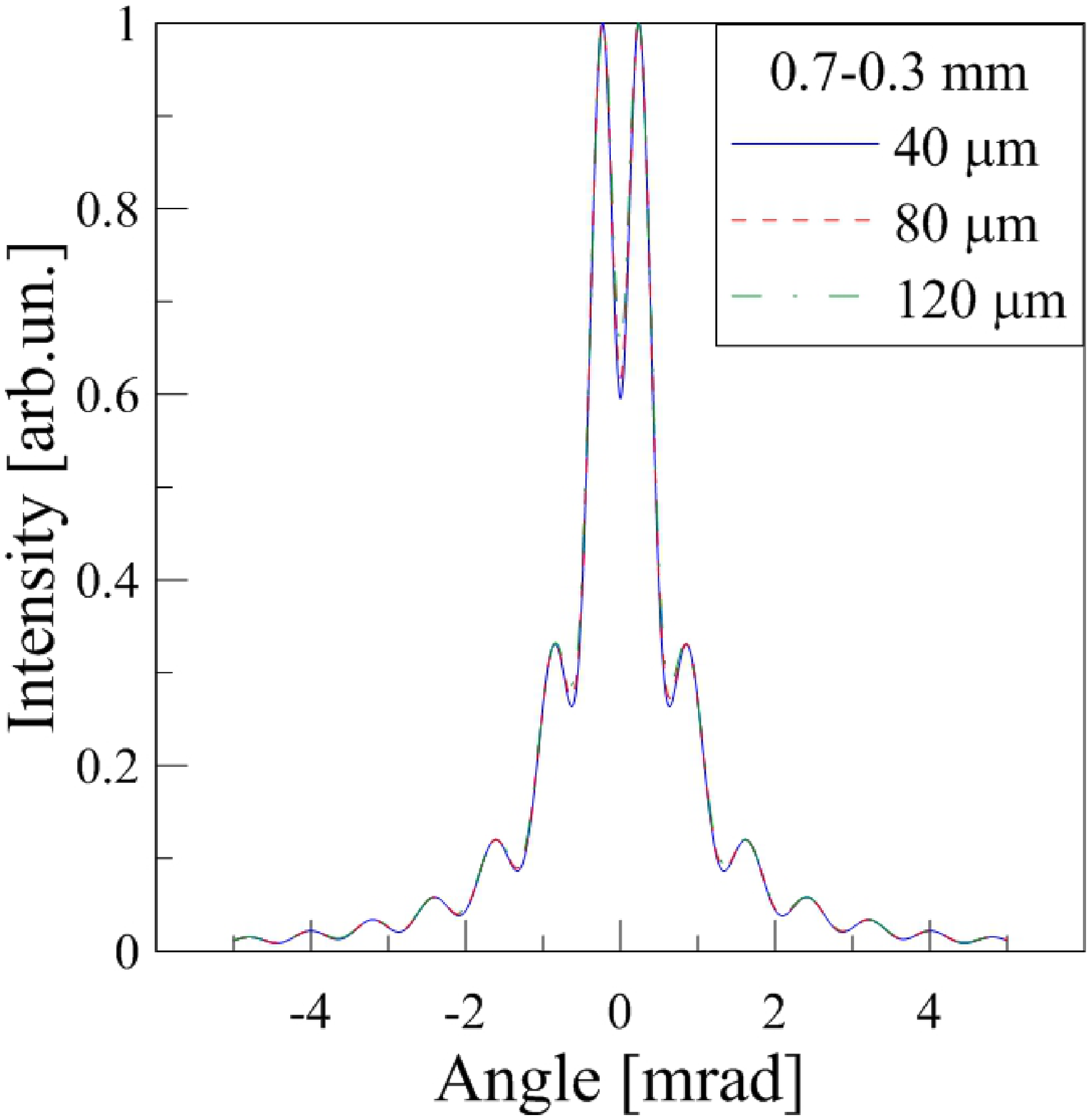}}
             ~
  \caption{Influence of the beam size on the DR distribution. On top of plots the first two values show the distances from both upper and lower half planes, and for the each plot is shown corresponding beam size.}
  \label{fig5}
\end{figure}
For our calculations the DESY parameters were used. In particular, the beam size was about 80 $\mu$m. In Fig.\ref{fig5} on top of plots the first two values, as in previous case, show the distances from both upper and lower half planes, and for the each plot is shown corresponding beam size. As seen from the plots in Fig.~\ref{fig5} the beam size influences on the distribution of DR but without any spikes' shift.

\subsection{Beam divergence}
Now, knowing that the beam size has no influence on the position of spikes, we can introduce the divergence of a beam in our calculations. According to \cite{6,7} in order to calculate the distribution of DR from a divergent beam we have to take the convolution of the intensity function with the function of angular distribution taken in Gaussian approximation. Then, we obtain
\begin{equation}
I_c=\int_c^d \! I_f(\phi-\Delta_\phi)F(\Delta_\phi)d\Delta_\phi\,\, ,
\end{equation}
or
\begin{equation}
I_c=\int_c^d \! \left[\int_a^b \! |E_2(y_0, \phi-\Delta_\phi)|^2G(y_0)dy_0\right]F(\Delta_\phi)d\Delta_\phi\,\, ,
\end{equation}
where $c$ and $d$ are the limits describing the effective divergence of a beam, and $\Delta_\phi$ is the deviation angle with respect to the normal of a screen.
\begin{figure}
    \vspace{-2cm}
  \centering
  \subfloat[]{\includegraphics[width=0.45\textwidth]{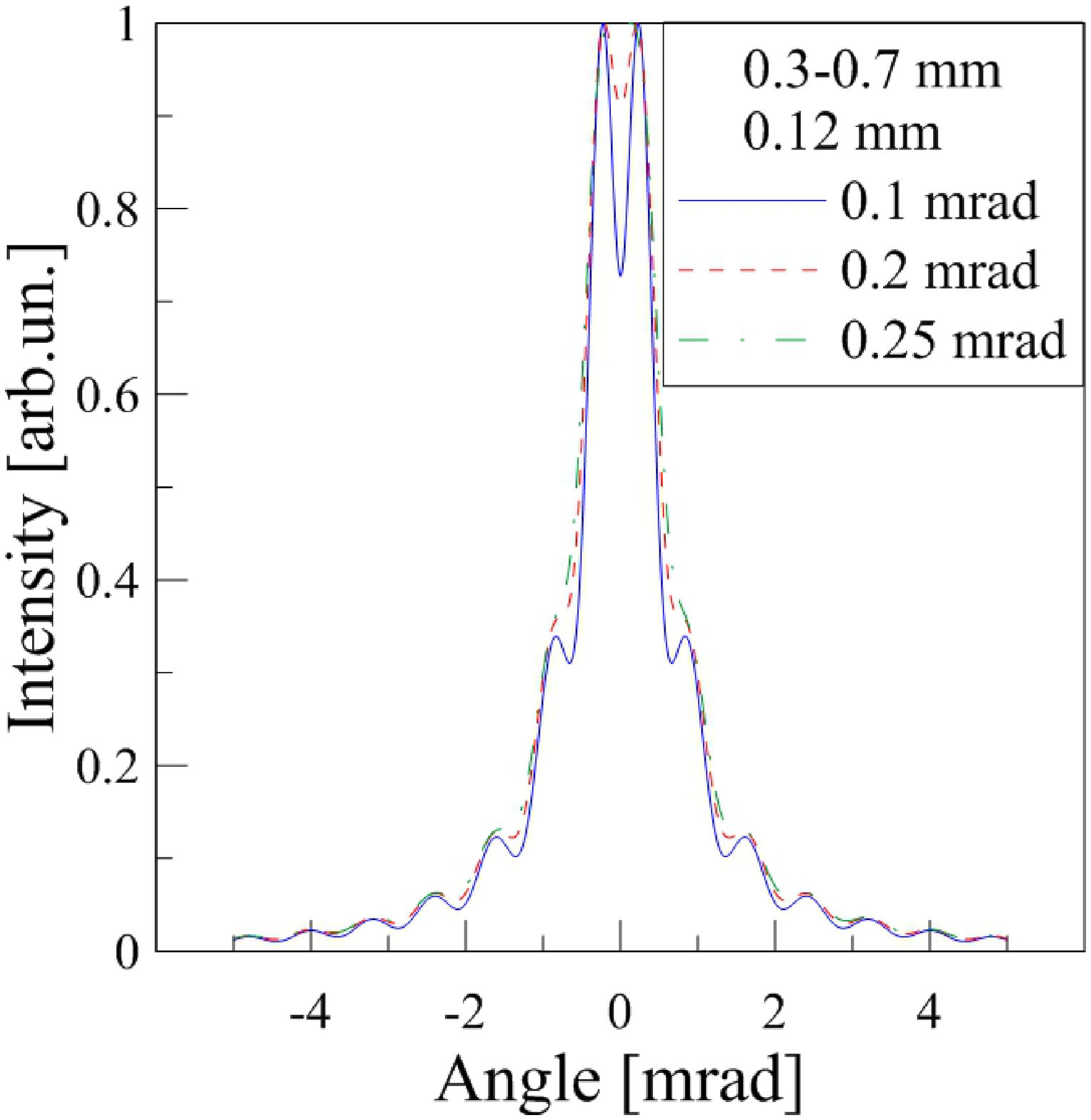}}
               ~
  \subfloat[]{\includegraphics[width=0.45\textwidth]{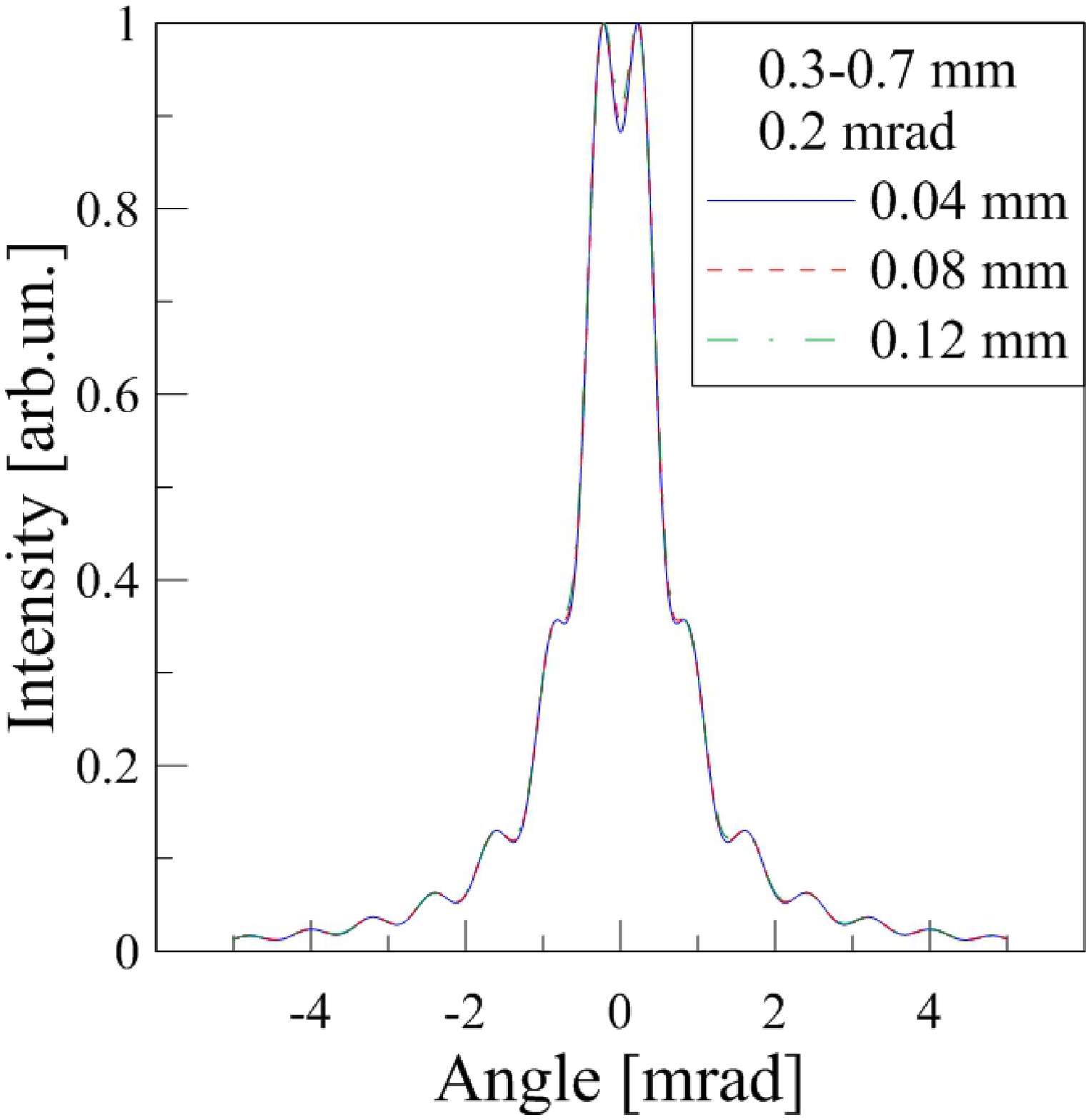}} \\
               ~
  \subfloat[]{\includegraphics[width=0.45\textwidth]{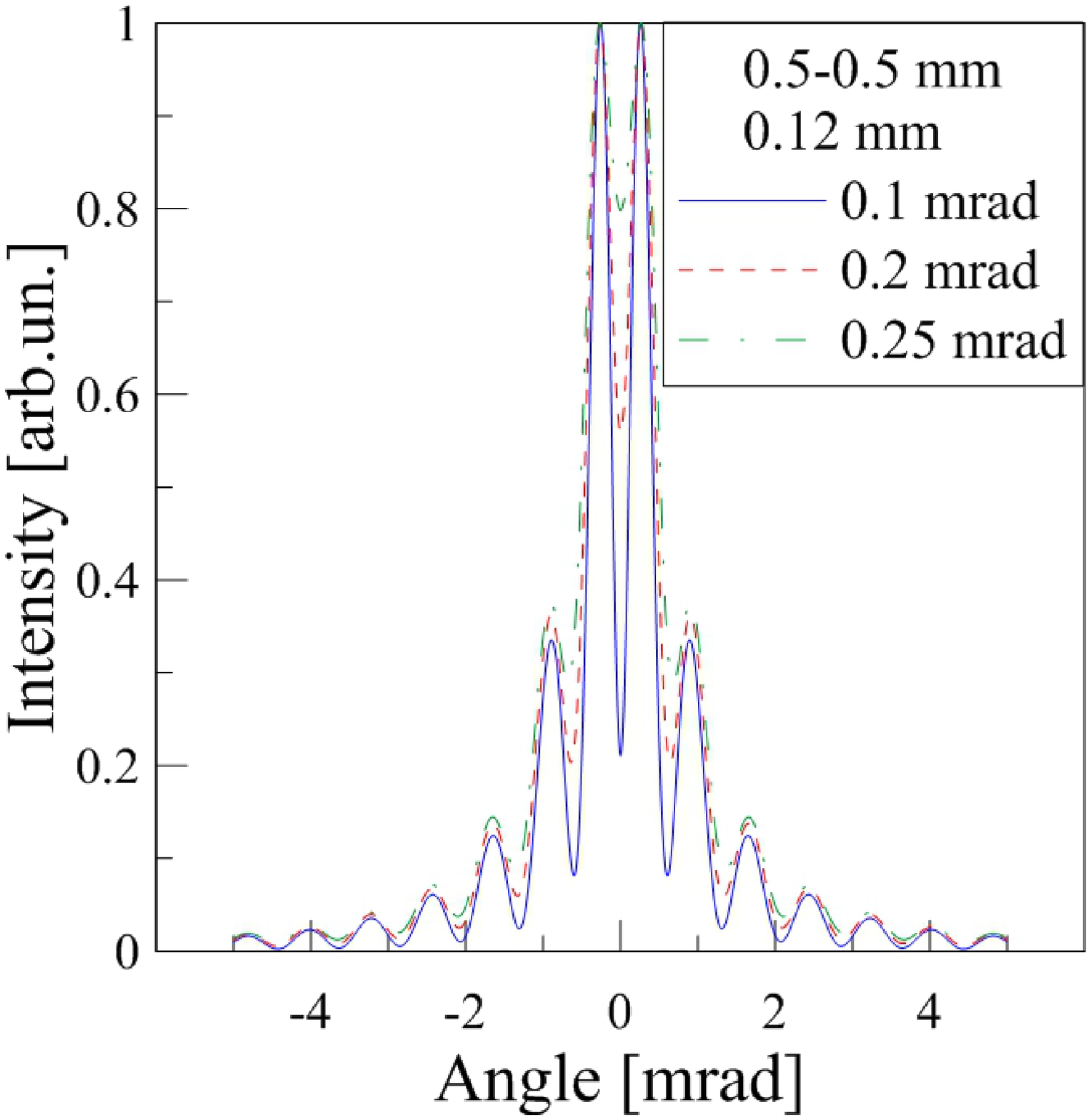}}
               ~
  \subfloat[]{\includegraphics[width=0.45\textwidth]{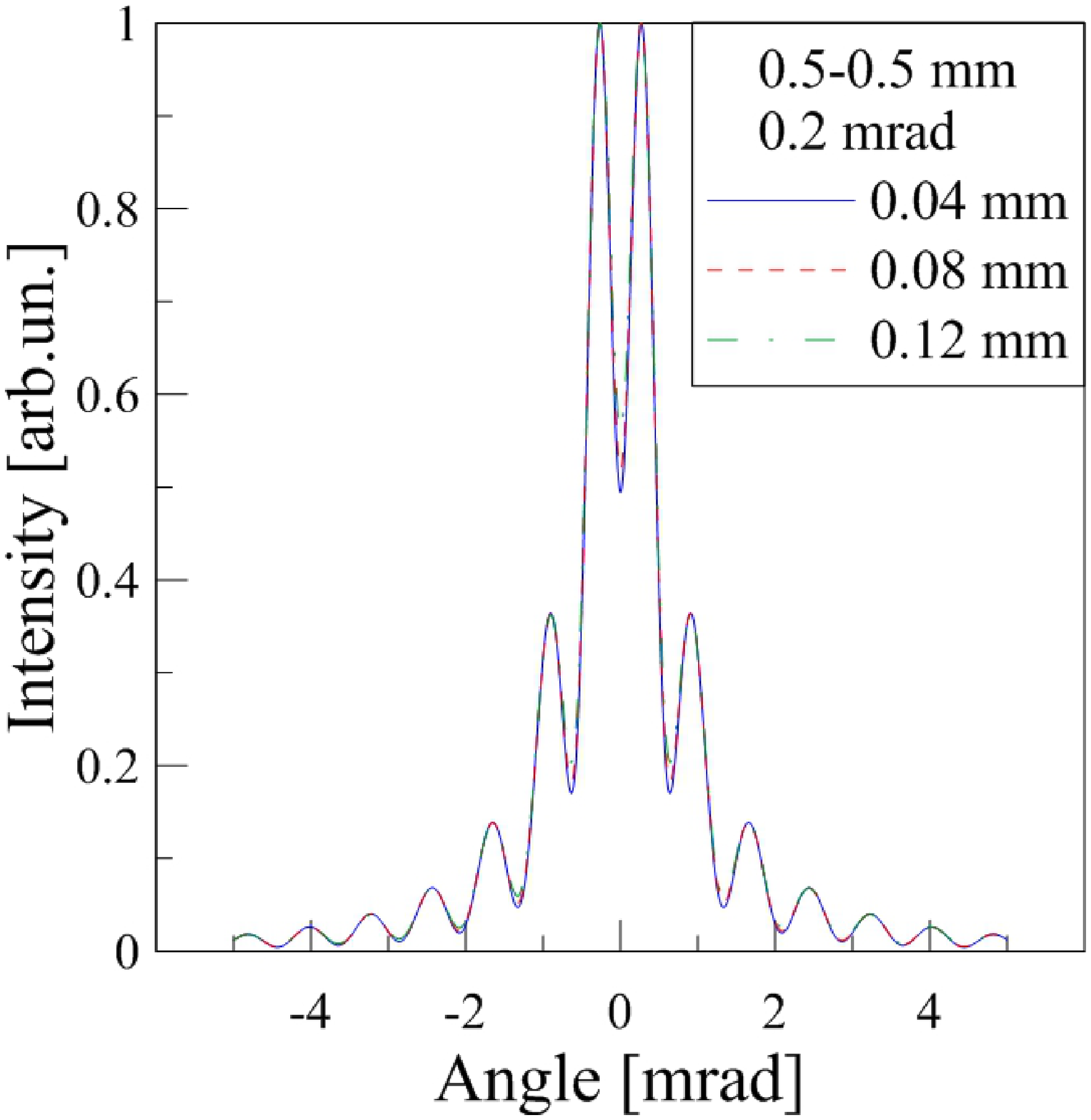}} \\
               ~
  \subfloat[]{\includegraphics[width=0.45\textwidth]{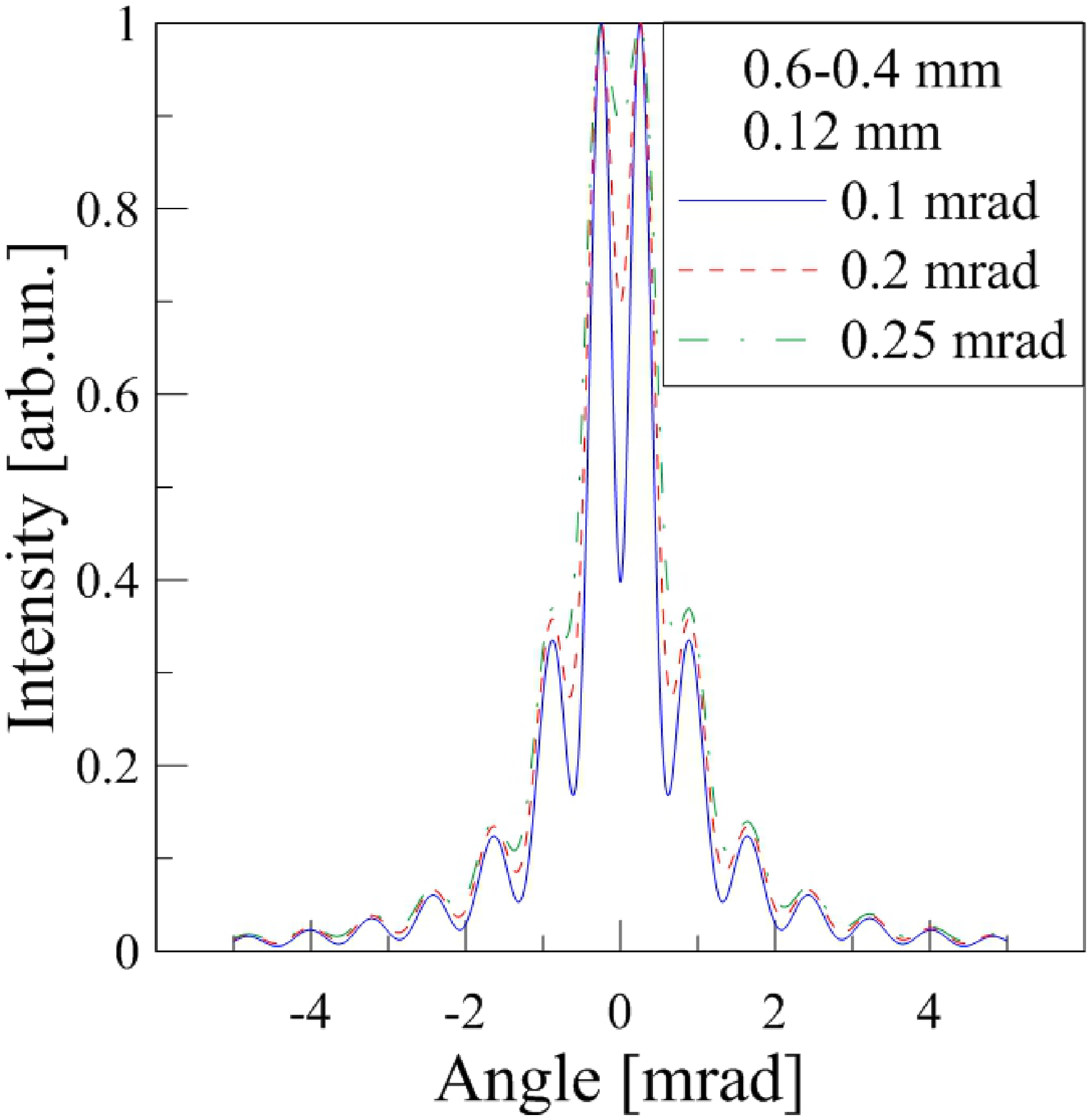}}
               ~
  \subfloat[]{\includegraphics[width=0.45\textwidth]{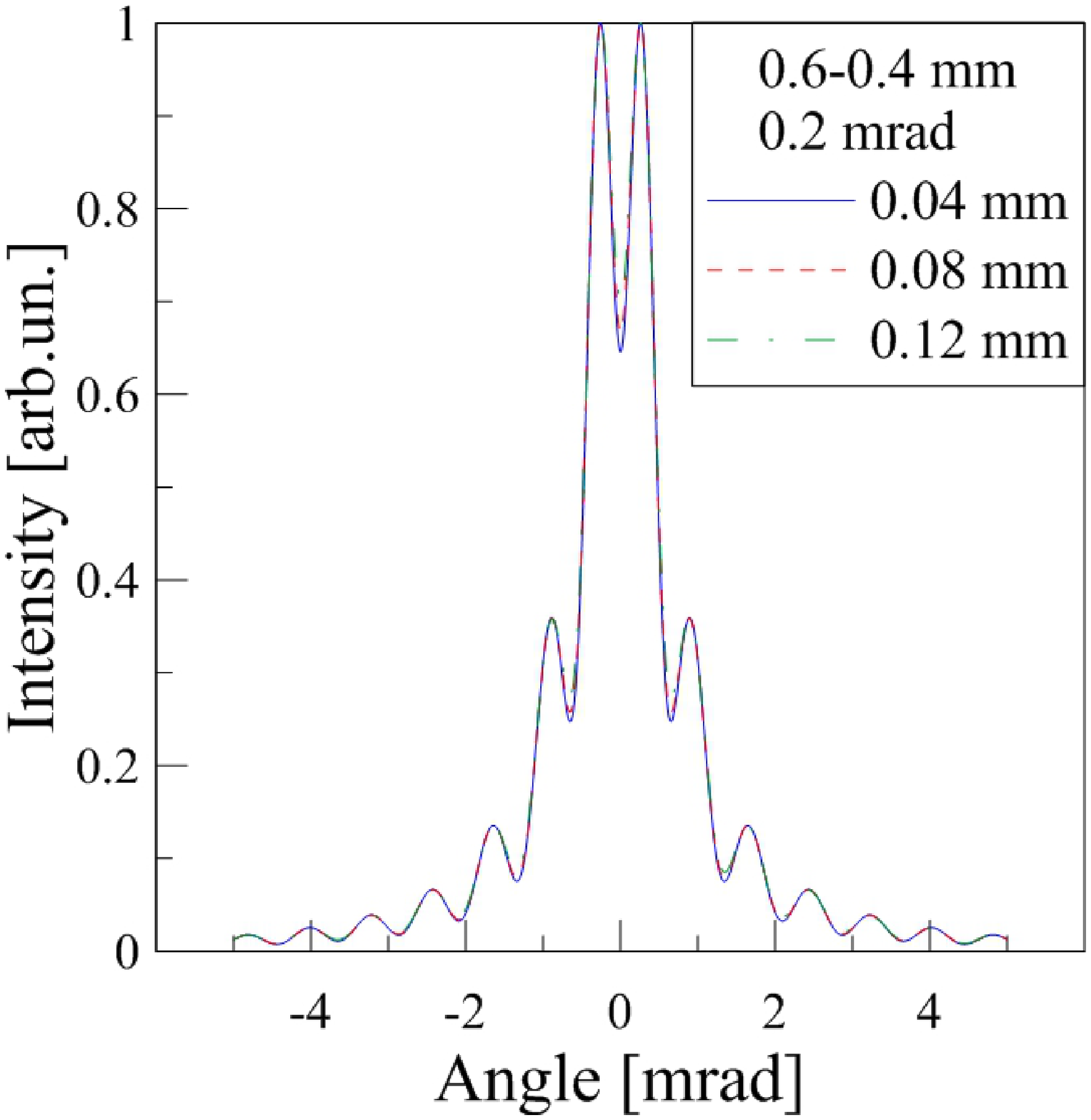}}
               ~
  \caption{The divergence dependence for the DR angular distribution. In right column first two values show the distance from beam line to upper and lower half-plane respectively, third value show the beam size, and each curve corresponds to the average beam divergence. In left column third value show the beam divergence, and each curve corresponds to the average beam size.}
  \label{fig6}
\end{figure}
In Fig.~\ref{fig6} (right column) you can see the influence of beam divergence on the DR distribution. First two values show the distance from beam line to upper and lower half-plane respectively, third value show the beam size, and each curve corresponds to the average beam divergence (in our case the FWHM, i.e. the full width at half maximum). The left column shows the distributions for different beam sizes but with the same divergence angle. As we can see the influence of angular divergence is much stronger than that of the beam size. However, this depends on the  parameters used.

As seen from this chapter, the divergence as well as the size of a beam have no influence on the position of spikes in the DR angular distribution. Unfortunately, these parameters might  lead to disappearing all spikes on the graphs, if they are comparable with the slits' values. This feature is very sharp in Fig.~\ref{fig6}, the right column.
\newpage
\section{System of two slits}
\begin{figure}[H]
    \begin{minipage}{0.4\textwidth}
      \includegraphics[scale=0.4]{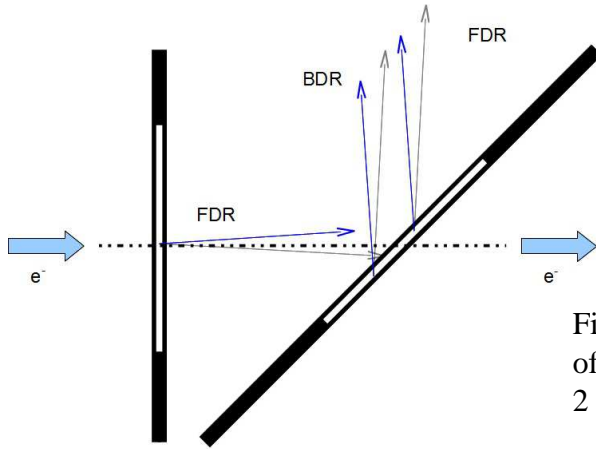}
      \end{minipage}
      \hfill
      \begin{minipage}{0.5\textwidth}
      \vspace{4cm}
      \caption{Two slit system, top view. Width of the first slit is equal to 1 mm,  the second - 2 mm.}
      \label{fig7}
      \end{minipage}
\end{figure}

In order to have more sensitive system based on DR technique let us introduce a second slit (Fig.\ref{fig7}).
For two slit configuration one of the most important parameter is the shift between the upper and lower half-planes, because even small shift between half-planes can lead to dramatic changes in DR distribution, as shown in \cite{7}. Moreover, the shift between the centers of slits should be taken into account. First of all, let us define the angular dependence between two spikes versus position of a bunch. For simplicity, we have chosen the slits to be collinear. To calculate DR from the system of two slits we have used the approximation as described in \cite{9}
\begin{figure}[H]
%\vspace{}
    \begin{minipage}{0.35\textwidth}
      \includegraphics[scale=0.4]{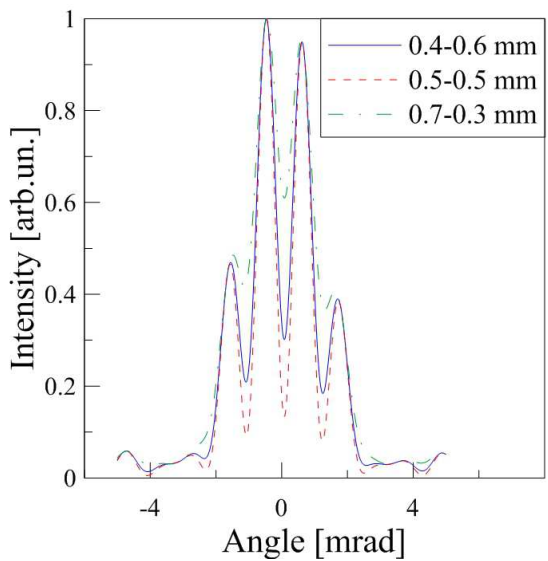}
      \end{minipage}
      \hfill
      \begin{minipage}{0.4\textwidth}
      \vspace{5.5cm}
      \caption{Dependence of DR on position of a bunch relative to the center of a slit.}
      \label{fig8}
      \end{minipage}
\end{figure}

In Fig.~\ref{fig8} we have represented several distributions corresponding to different locations of the beam. As we can see, for a two-slit configuration, the distance between spikes does not vary at the change of electron position that makes the system of two slits not applicable for determination of electron bunch position.

\section{Near-field calculations}

In previous sections for all calculations a far-field approximation was used. However, in  experiment with two-slit setup the second slit is situated in 2 cm from the first one. Hence, in order to know the changes in final DR distribution due to the presence of a second slit  we need to know DR distribution from the first slit at the position of second one. These calculations have to be done within the near-field approximation, which we have used as for transition radiation (TR) \cite{10}:

\begin{multline}
E_{x,y}^{TR}(a, c)=-\frac{k}{d4\pi^2}\int_{-a+h_{0y}}^{a+h_{0y}} \! dh_y\int_{-c+h_{0x}}^{c+h_{0x}} \! dh_xE_{x,y}^i(h_x,h_y) \times \\
\\ \times \frac{\exp{[ik\sqrt{d^2+(h_x-x)^2+(h_y-y)^2}]}}{\sqrt{d^2+(h_x-x)^2+(h_y-y)^2}}\,,
\label{E-tr}
\end{multline}
where
\begin{equation}
E_{x,y}^i(h_x,h_y)=\frac{ek}{\pi\gamma}\frac{h_x,h_y}{\sqrt{h_x^2+h_y^2}}K_1\left(\frac{k}{\gamma}\sqrt{h_x^2+h_y^2}\right)
\end{equation}
\begin{figure}[H]
\vspace{-0.5cm}
    \begin{minipage}{0.5\textwidth}
      \includegraphics[scale=0.35]{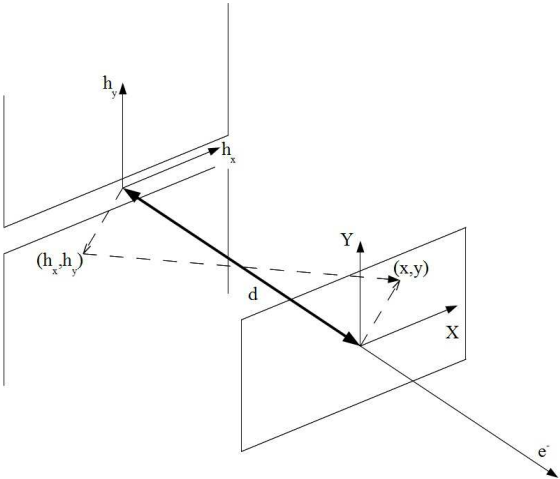}
      \end{minipage}
      \hfill
      \begin{minipage}{0.4\textwidth}
      \vspace{5.5cm}
      \caption{Sketch for near-field calculation.}
      \label{fig9}
      \end{minipage}
\end{figure}
Here $K_1$ is the McDonald function, $a$ and $c$ are the sizes of a plate, $h_{0y}$ and $h_{0x}$ are the initial shifts with respect to the center of a plate, and $d$ is the distance between the TR source and the plane of radiation detection (Fig.\ref{fig9}). In order to calculate DR, we have to exclude from the integral expression for TR the part corresponding to the slit. Indeed, it means that we can consider DR as TR from two half-planes without the coefficient before the integral (\ref{E-tr})
\begin{multline}
\label{near-field}
E_y^{DR}(a,c)=\int_{\frac{w_{sl}}{2}-h_{0y}}^{a} \! dh_y \int_{-c}^{c}E_{y}^i(h_x,h_y) \frac{\exp{[ik\sqrt{d^2+(h_x-x)^2+(h_y-y)^2}]}}{\sqrt{d^2+(h_x-x)^2+(h_y-y)^2}} dh_x +\\
\\ +\int_{-a}^{\frac{-w_{sl}}{2}-h_{0y}} \! dh_y \int_{-c}^{c}E_{y}^i(h_x,h_y) \frac{\exp{[ik\sqrt{d^2+(h_x-x)^2+(h_y-y)^2}]}}{\sqrt{d^2+(h_x-x)^2+(h_y-y)^2}} dh_x,
\end{multline}
where $w_{sl}$ is the width of a slit.

Now let us analyze and successfully compare the results of calculations for both far- and near-field approximations.

\begin{figure}[H]
  \centering
  \subfloat[]{\includegraphics[width=0.45\textwidth]{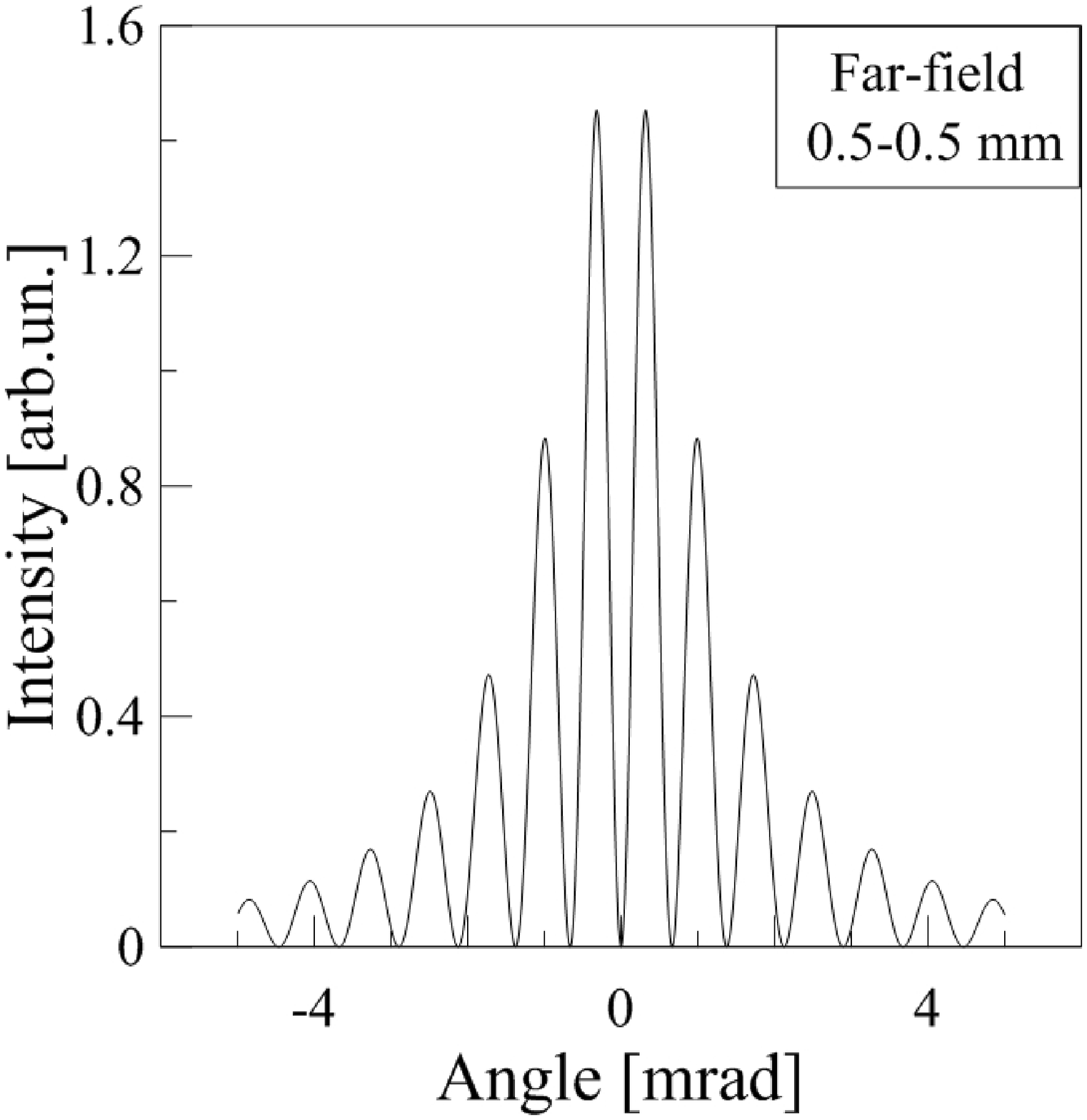}}
  ~
  \subfloat[]{\includegraphics[width=0.45\textwidth]{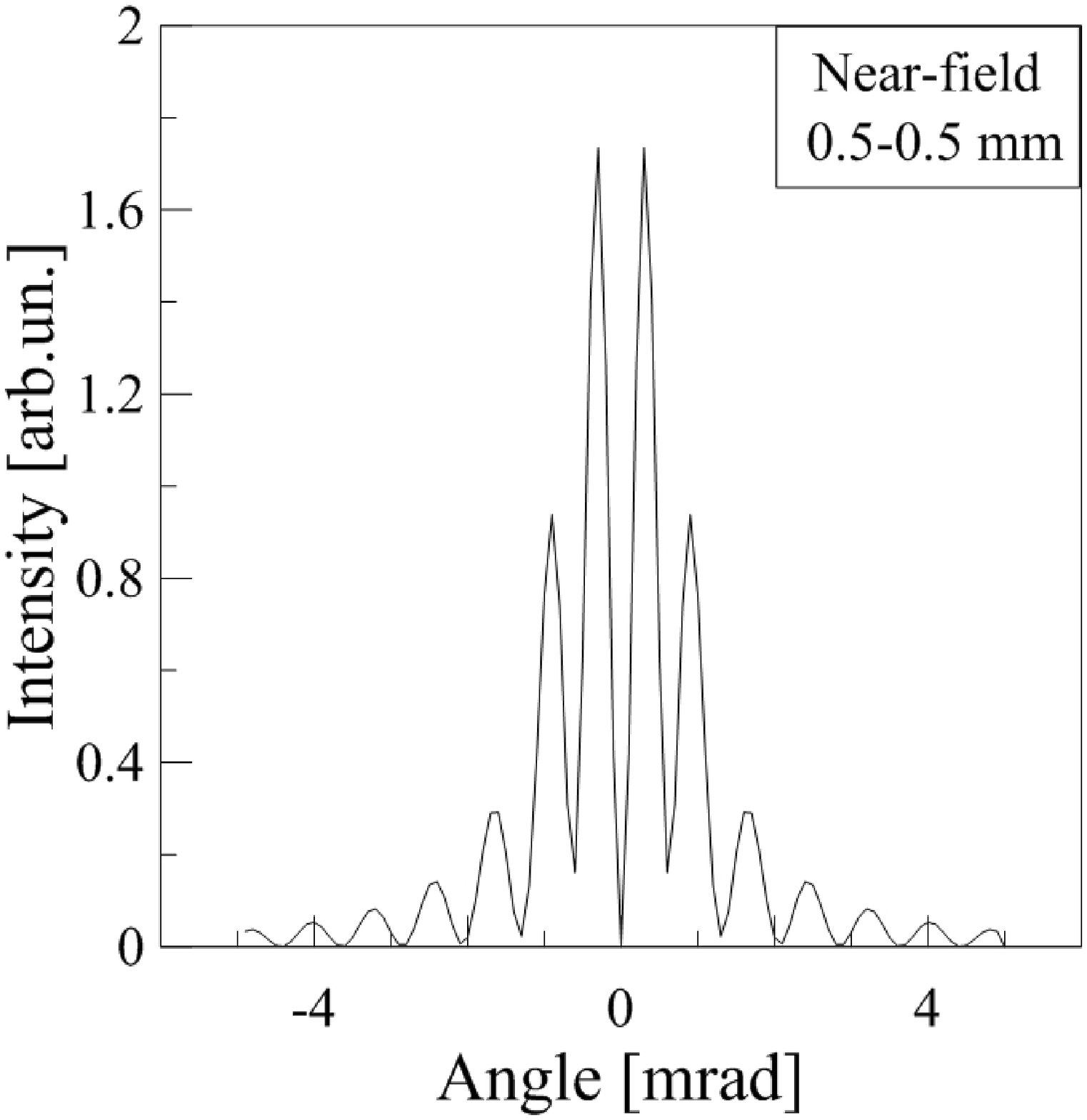}} \\
  ~
  \subfloat[]{\includegraphics[width=0.45\textwidth]{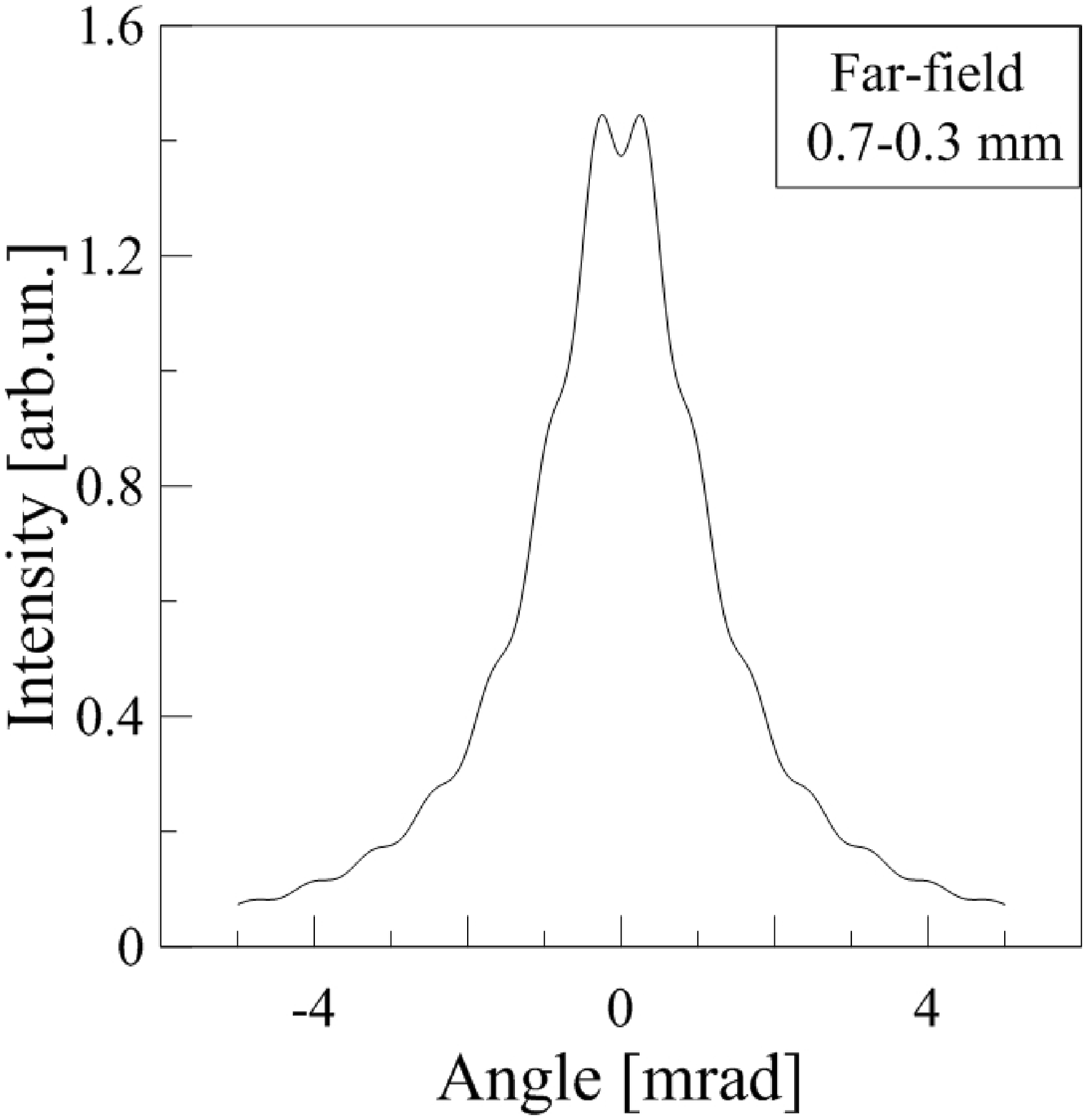}}
  ~
  \subfloat[]{\includegraphics[width=0.45\textwidth]{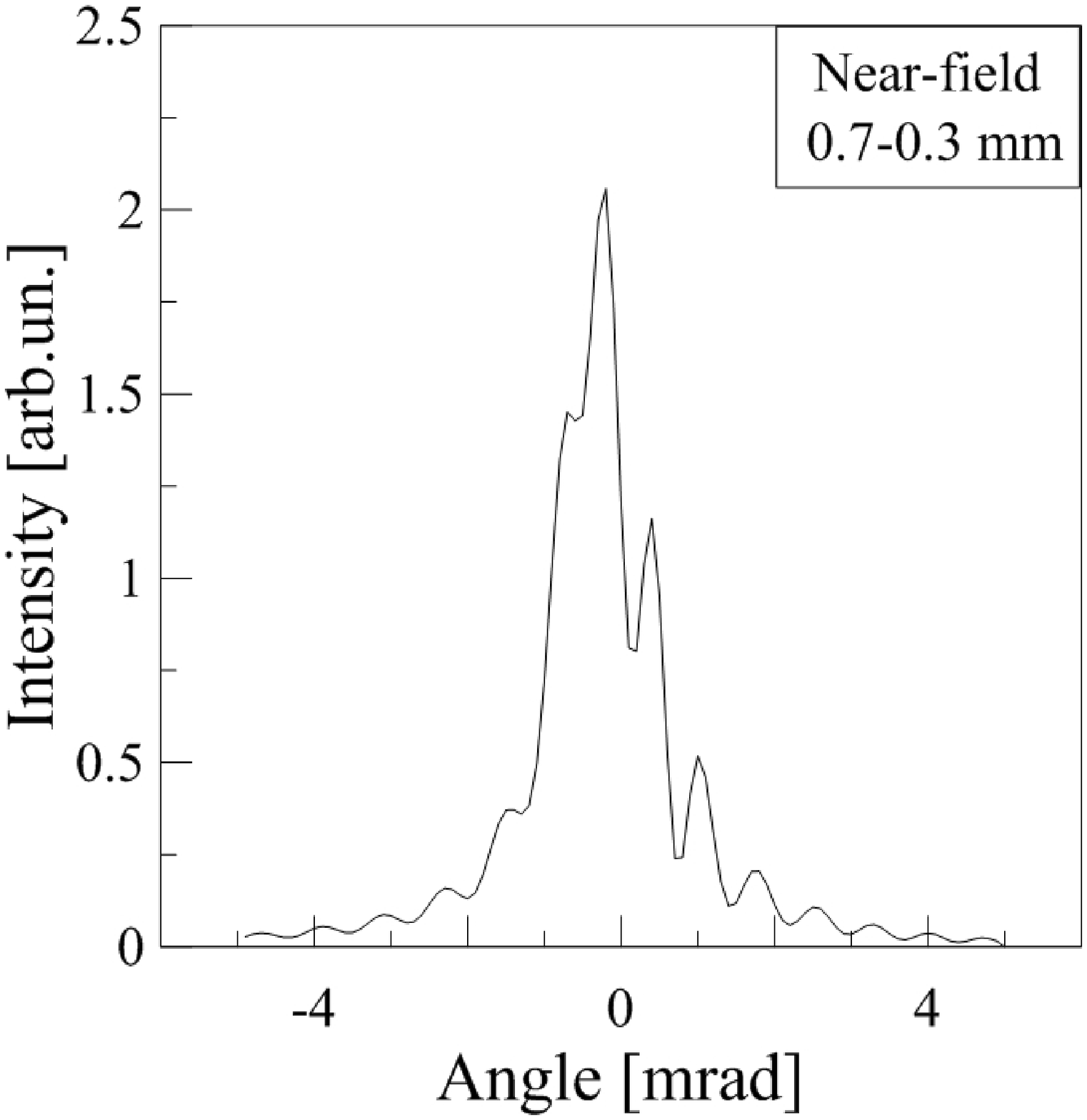}}
  ~
  \caption{Far- and near-field approximations for DR.}
  \label{fig10}
\end{figure}
 In Fig.~\ref{fig10} a) and b) we have presented the angular distributions for both far- and near-field calculations when electron passes through the center of a slit with the width of 1 mm. As seen, the distributions are characterized by visible differences, while the positions of maxima remains unchanged. We deal with the opposite situation when the bunch is shifted with respect to the center of a slit. Namely, the DR angular distributions simulated for far-field approximation are different from those for near-field one. This situation is shown in Fig.~\ref{fig10} c) and d) (here, the impact parameter for upper half plane is 0.7 mm, and for lower half plane - 0.3 mm). This feature becomes easy to understand if we consider DR from one half-plane.
\begin{figure}[H]
\vspace{-1cm}
  \centering
  \subfloat[]{\includegraphics[width=0.47\textwidth]{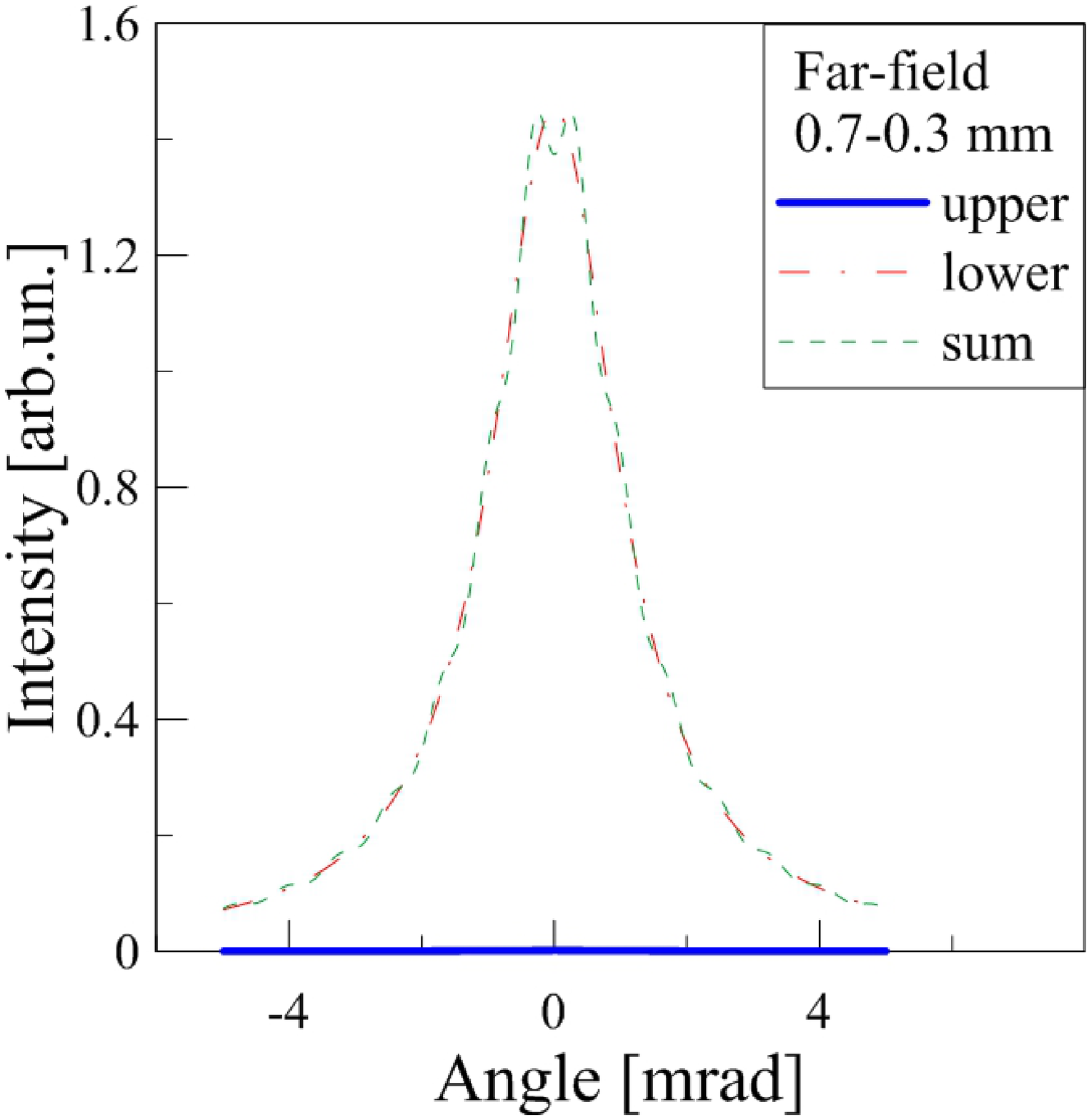}}
  ~
  \subfloat[]{\includegraphics[width=0.47\textwidth]{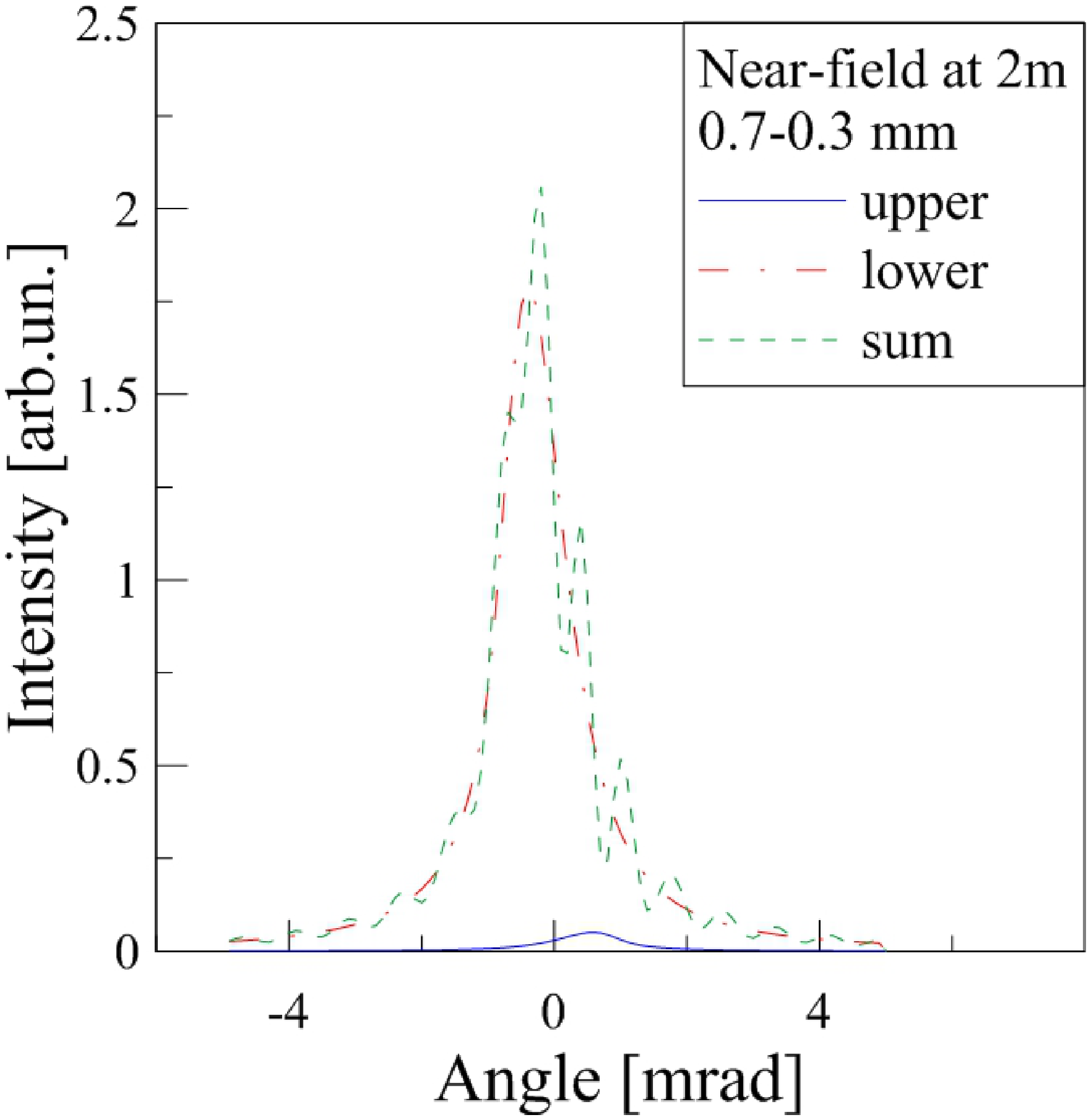}}
  ~
  \caption{DR from half-planes of one slit in far- and near-field approximations.}
  \label{fig11}
\end{figure}

 In Fig.~\ref{fig11} for both pictures the blues line represent the distributions from the upper half-plane, the red line - from lower half-plane, and the dashed line - the sum of these distributions. From the distributions the intensity maxima in far-field approximation are revealed in forward direction (at zero angle), while for near-field distributions the maxima are shifted from "zero" in opposite directions to each other.

\begin{figure}[H]
  \begin{minipage}{0.5\textwidth}
  \includegraphics[scale=0.37]{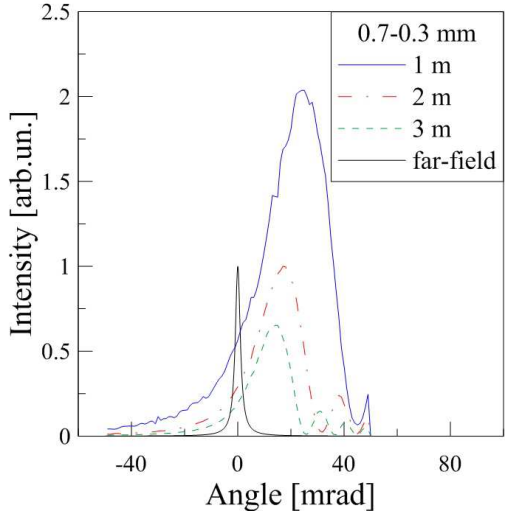}
  \end{minipage}
  \begin{minipage}{0.47\textwidth}
  \hfill
  \vspace{2.2cm}
  \caption{DR intensity distributions vs position of the detection plane. The blue solid curve corresponds to 1 m distance, red dash-dot curve - 2 m, green dased curve - 3 m. Black  thin solid curve, which  corresponds to the far-field calculations, has been normalized to be evident; in reality this intensity is much low.}
  \label{fig12}
  \end{minipage}
\end{figure}

In Fig.~\ref{fig12} we can see that the position of maximum for DR angular distribution, simulated in near-field limit, tends to zero at moving the detection plane far away from the slit that corresponds to the far-field approximation at long distances. As seen, the DR intensity at such long distances are extremely low that is, obviously, expected result (it is shown by normalized black curve). Moreover, from these calculations we can conclude that the pronounced maximum of DR in near-field case becomes less sharp by its intensity for a distant observation plane.

It should be underlined that in experiments the angular distributions are measured by mean of a lens that allows direct (distance independent) angular scan for DR distributions. This feature is taken into account for further simulations.

\section{Shift between the centers of two slits}
Let us apply near-field calculations to two-slit system. Typically, we measure the interference between forward DR (FDR) from the first slit and backward DR (BDR) from the second slit (Fig.\ref{fig7}). At reflection of FDR from the second slit some portion of radiation passes through the hole; evidently, this portion does not contribute to BDR. In order to calculate this part we have to use the near-field calculations. Spatial distribution of FDR at the distance of 2 cm for 1 mm slit can be simulated by the expression (\ref{near-field}).
\begin{figure}[H]
  \begin{minipage}{0.5\textwidth}
  \includegraphics[scale=0.35]{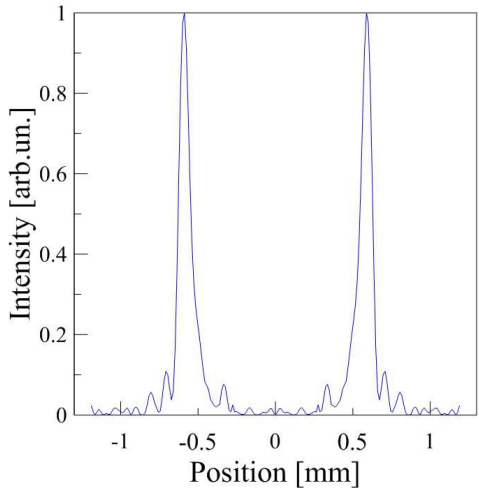}
  \end{minipage}
  \begin{minipage}{0.45\textwidth}
  \vspace{4cm}
  \hfill
  \caption{FDR at the second slit position.}
  \label{fig13}
  \end{minipage}
\end{figure}
At such small distances total DR, which is generated by two half planes, does not still overlap. Thus, at the second slit position we can resolve two independent maxima. Indeed, the emission distributions overlap in the hole space of a slit that results in its loosing for further evaluation. However, this portion of radiation is rather small (the second slit of only 0.5 mm), and successfully, we can see that it does not change essentially the final distribution.

\begin{figure}[h]
    \begin{minipage}{0.5\textwidth}
    \includegraphics[scale=0.35]{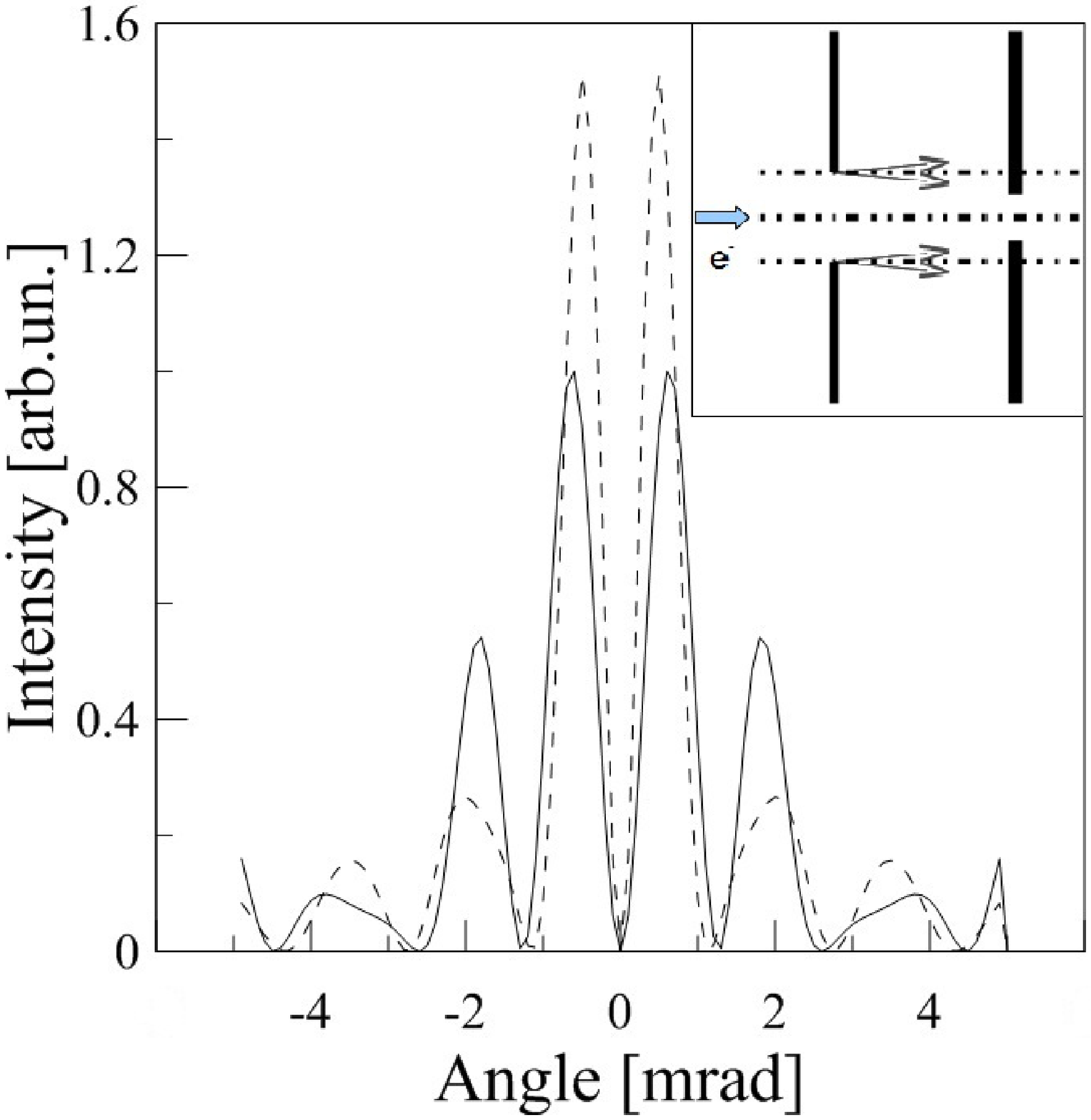} \\ \centering a)
    \end{minipage}\hfill
    \begin{minipage}{0.5\textwidth}
    \includegraphics[scale=0.35]{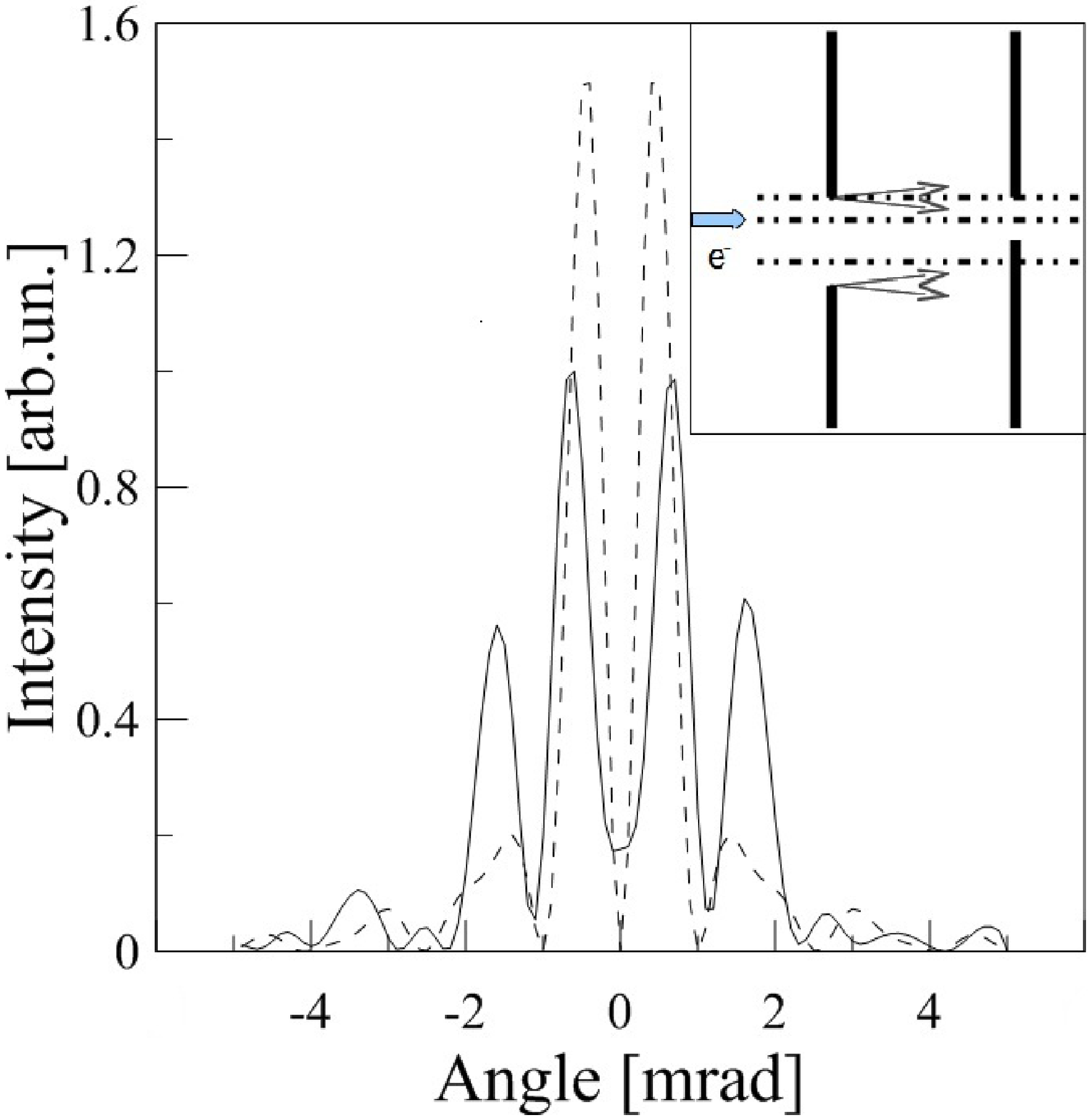} \\ \centering b)
    \end{minipage}\hfill
    \begin{minipage}{0.5\textwidth}
    \includegraphics[scale=0.35]{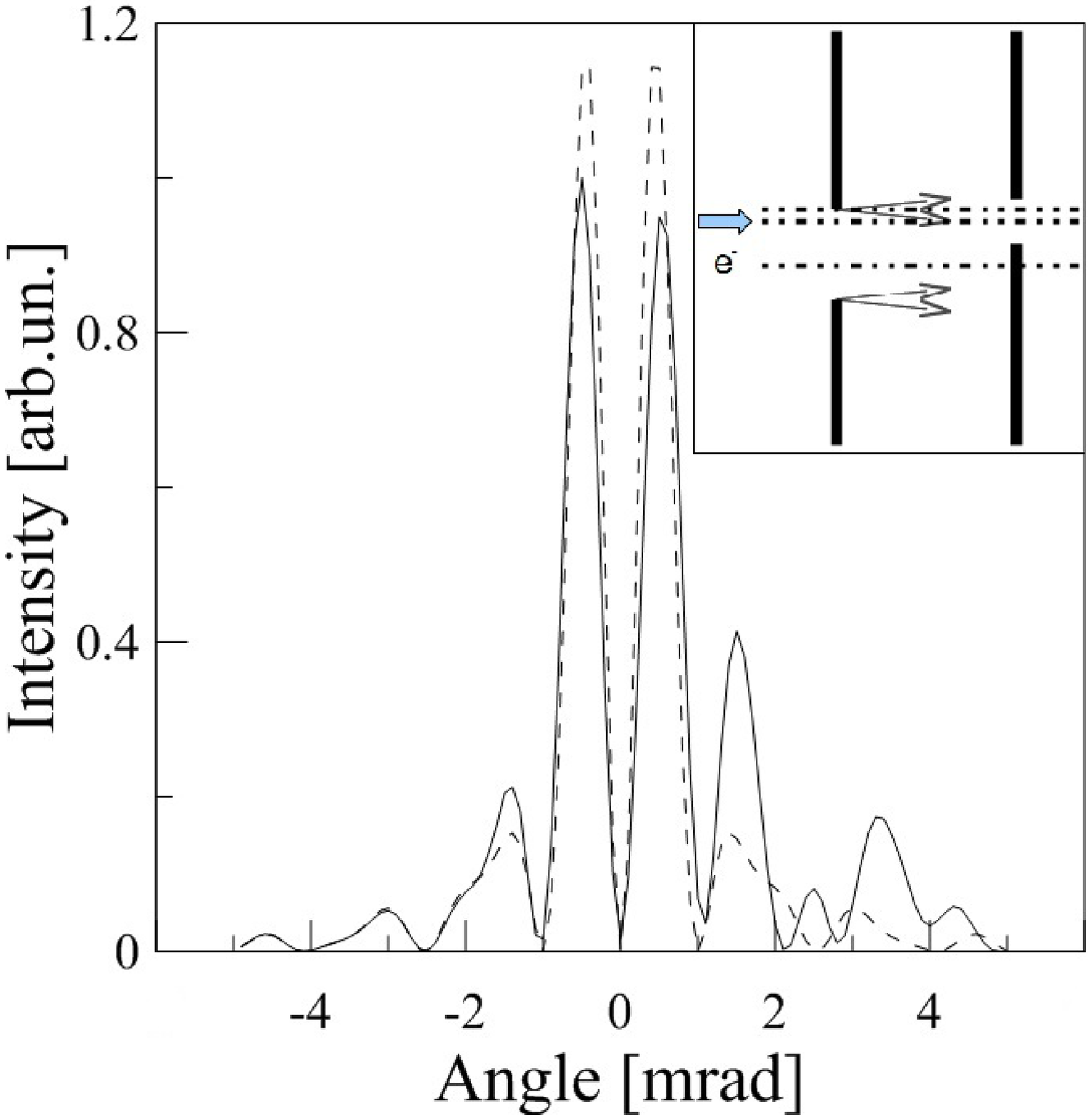} \\ \centering c)
    \end{minipage}
    \begin{minipage}{0.45\textwidth}
    \vspace{5cm}
    \caption{DR distribution for different positions of the slits.}
    \label{fig14}
    \end{minipage}
    \hfill
\end{figure}

In order to understand how DR distribution depends on the shift between the centers of two slits several angular distributions for various shifts were compared (Fig.~\ref{fig14}). First distribution (a) corresponds to the beam passing through the center of first slit without the slits shift. Dashed line is DR from the second slit in absence of the first one. In presence of the first slit the central maxima get lower and, vice versa, the second (side) maxima become higher.
Following the DESY experiments, the second slit is fixed to have a beam passed through its center, while the position of the first slit has been changed. In Fig.~\ref{fig14} b) we have shown the case when the edges of the upper half-planes of both first and second slits are on the same level. For such configuration we lose part of radiation from upper half-plane that corresponds to negative angles. As a consequence, the relative influence of the first slit on positive part of DR will be stronger. In Fig.~\ref{fig14} b) we observe the positive spike less intense than negative one (with respect to zero position), and, vice versa, for the second maxima. As above mentioned, for the collinear slits, additional slit makes central maxima less, while for the side maxima we have got the intensity increase.
Now let us decrease the impact parameter for upper half plane of the first slit with respect to that for the second slit (Fig.~\ref{fig14} c). For this geometry we can neglect the  radiation from lower half-plane of the first slit due to relatively distant beam pass. Radiation in negative angles (especially for big angles) is completely lost and, hence, its influence becomes disappearing. But some part of radiation for positive angles even more intensive than the radiation from the second slit, is reflected (Fig.~\ref{fig14} c). Comparing these results we can conclude that the influence of the slits shift is mostly significant when beam passes very close to one of the half-planes.

\section{Comparing with experiment}
In order to compare with experimental data it is necessary to introduce parameters of both beam and slit. For far-field calculations we have already introduced the size of a beam as all as its divergence. For near-field case, the calculations were performed paying attention to coherent features of a beam. Indeed in our calculations we deal with the field parameters instead of their intensities in the case of far-field limit, i.e. the coherent field can be written as follows

\begin{equation}
E_{coh}=\int_c^d \! E_2(y_0, \phi-\Delta_\phi)F(\Delta_\phi)d\Delta_\phi
\end{equation}

\noindent here $c$ and $d$ defines the beam divergence cone. Preliminary calculations have shown that the size of a beam has very small influence on final distributions. The latter allows us to limit the analysis taking into account the beam divergence. However, we have additionally to introduce the shift between two half-planes of the slit. This can be done by adding the phase shift between uppers and lowers half-planes of both slits \cite{9}. Moreover, in our calculations we have to introduce the angle of incident
\begin{figure}[h]
  \begin{minipage}{0.53\textwidth}
  \includegraphics[scale=0.35]{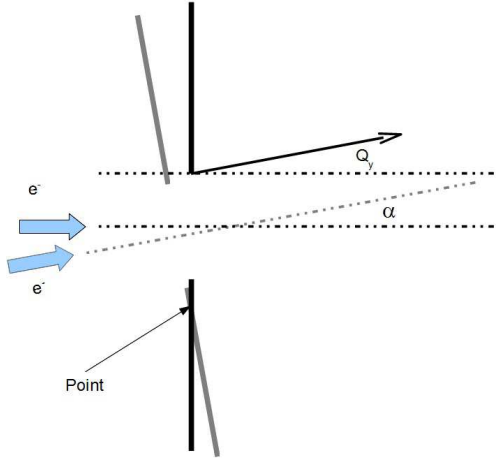}
  \end{minipage}
  \begin{minipage}{0.45\textwidth}
  \vspace{5cm}
  \caption{Schematical view for definition of the beam divergence.}
  \label{fig15}
  \end{minipage}
  \hfill
\end{figure}

 It is well known how to calculate DR for the geometry when the beam crosses the slit normal to the slit plane. To evaluate DR for the case of non perpendicular incidence we can  reduce the problem to known one. Namely, for each point of the slit we can apply imaginary slit perpendicular to the incident beam (Fig.~\ref{fig15}). That allows calculation of radiation within known code. However, within this approach it becomes necessary to change impact parameter for each point of the slit: $h'_l=h_l\cos\alpha$, as well as to take into account the phase shift between the points of a slit (we assume that projectile emits at shortest impact parameter).

\begin{figure}[h]
  \centering
  \subfloat[]{\includegraphics[width=0.4\textwidth]{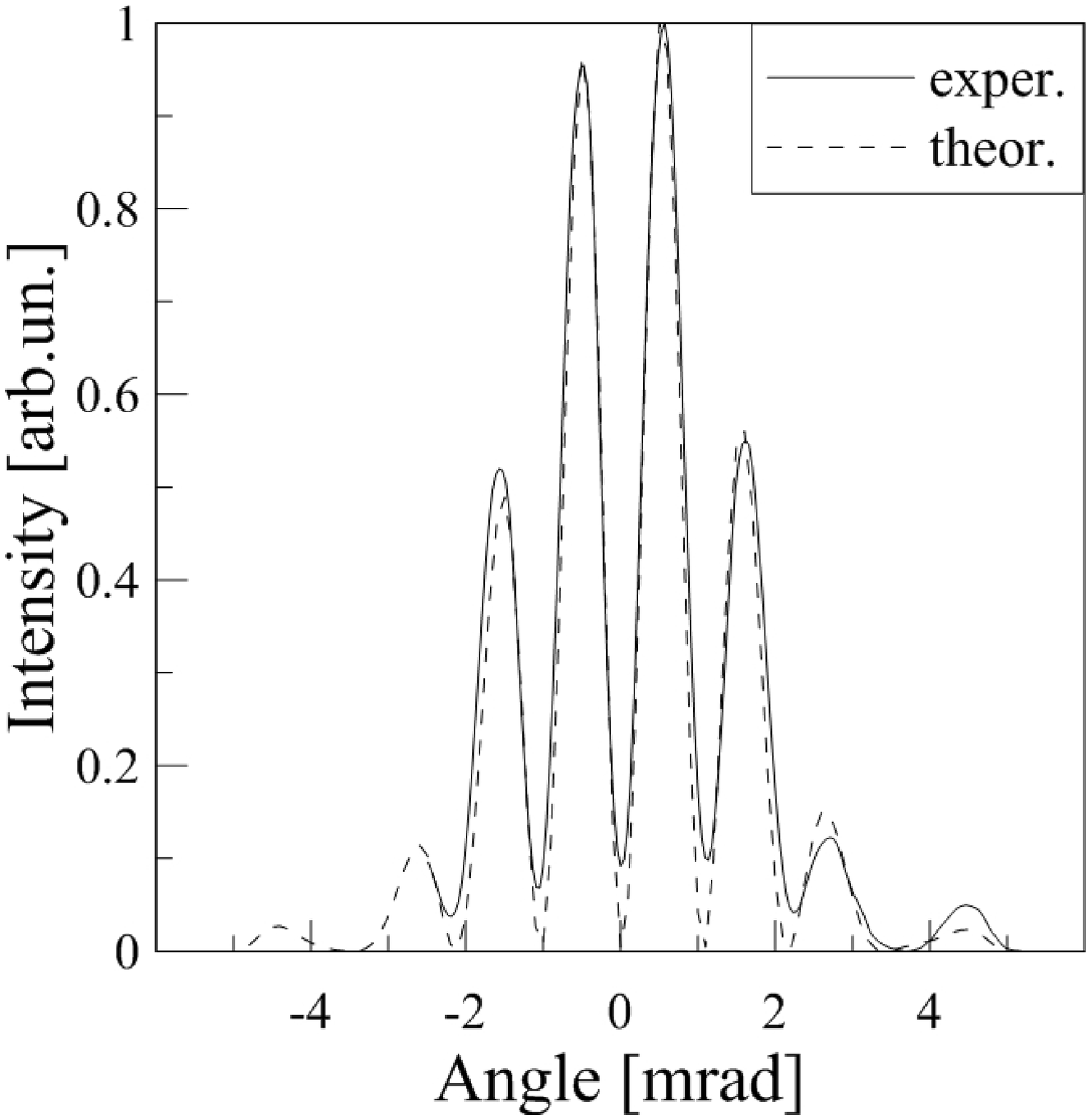}}
  ~
  \subfloat[]{\includegraphics[width=0.4\textwidth]{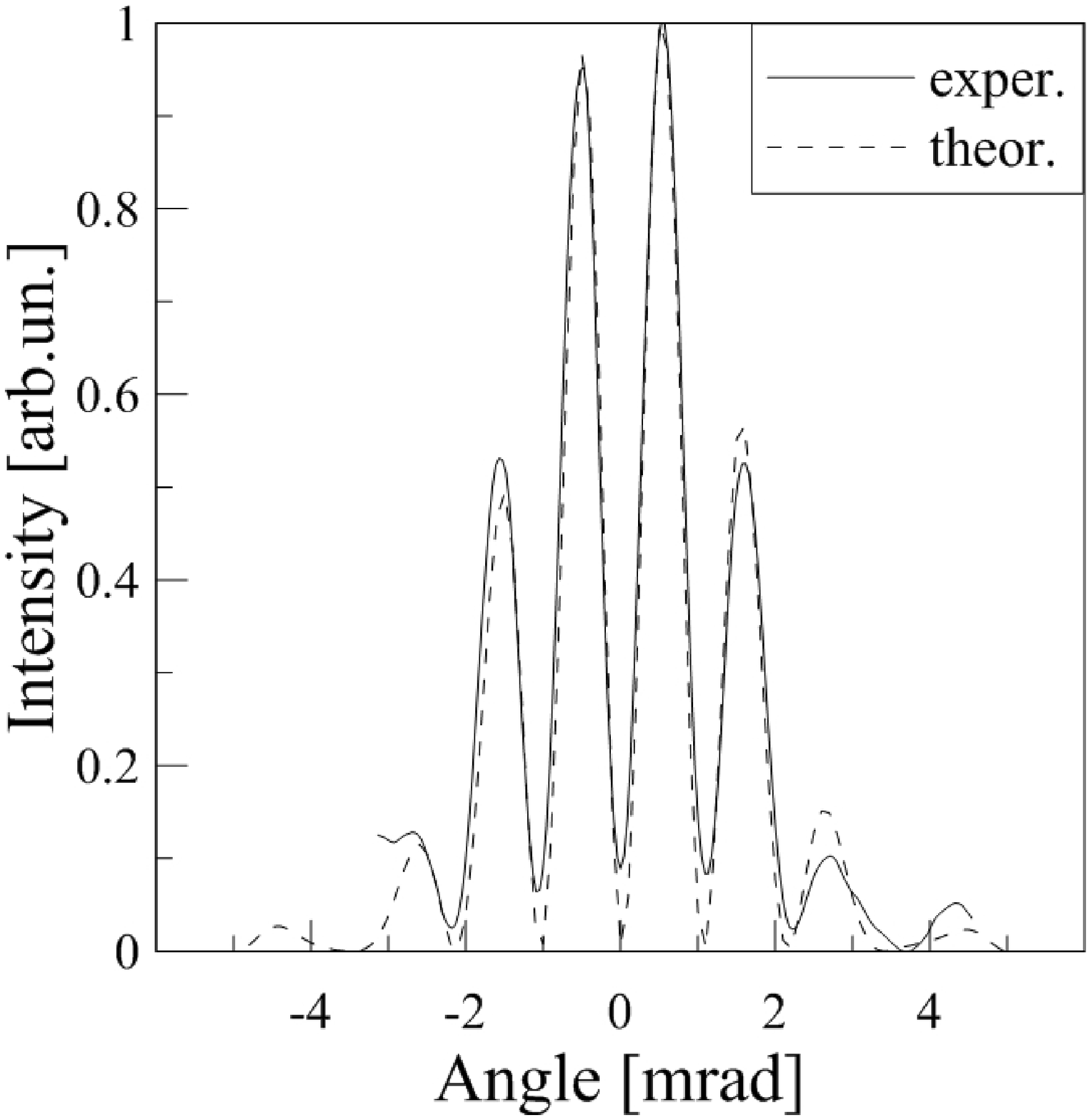}}
  ~
\caption{In experiment a second slit fixed and a beam passes at the center, while a first slit is moveable. a) the beam passes both slits at the center; b) first slit is shifted by 50 $\mu$m.}
\label{fig16}
\end{figure}

The experimental DR distributions are presented in Fig.~\ref{fig16}. In experiments two-slit system was used with the distance between the silts of 2 cm; the slits width were 1 mm for the first slit, and 0.5 mm for the second. The electron beam passes through the center of second slit, while the first slit is moveable. The accuracy in half-planes coplanarity was within about 10 nm. The beam divergence was measured about 10 $\mu$rad.

Fig.\ref{fig16} a) shows DR when the beam passes the centers of both slits, while in the figure b) the first slit is shifted by 25 $\mu$m. On both graphs the dashed line corresponds to  theoretical dependences, while the solid line - to experiment ones.

As seen, theoretical distributions are in a good agreement with experimental data keeping the maximal error within 5\%. For calculations we have defined next parameters (close to those of the experiment): the shift between slit half-planes of 60 nm, the divergence of 40 $\mu$rad. Our calculations have shown better coincidence with experiment at the second slit width of 0.57 mm instead of 0.5 mm used.

\section*{Conclusion}

In this work we have presented the results on DR in near-flied limit, and its difference from the one of far-filed approach. Having calculated spatial distribution of DR in near-filed zone, we have studied the influence of slits incoplanarity on the final distribution. The results obtained have shown that at definite system configurations the relative shift of the slit centers can strongly affect the resulting DR angular distribution. The simulations were first done for the case of coherent beam. All described models were used to compare with experimental data, and have shown a good coincidence. Calculated beam parameters, such as its divergence, were well comparable with beam parameters obtained by the use of alternative methods (quadruple scanning).

Moreover, in this work we have proposed a new method to define the position of electron beam with respect to the slit taking into account the positions of two central maxima in DR angular distributions.
\section*{Acknowledgements}
The authors are grateful to M. Ferrario, M. Castellano, A. Cianchi and E. Chiadroni for fruitful discussions and experimental data provided. Special thanks from V. Sh. addressed to O. Bogdanov, A. Babaev and E. Frolov for their help in numerical calculations.
%\newpage

\end{document}